# Exploring structure diversity in atomic resolution microscopy with graph neural networks

*Zheng Luo[1,8], Ming Feng[2,3,8], Zijian Gao[2,8], Jinyang Yu[2], Liang Hu[3], Tao Wang[4], Shenao Xue[1,5], Shen Zhou[6], Fangping Ouyang[5], Dawei Feng[2], Kele Xu[2]\*& Shanshan Wang[1,7]\**

[1]Department of Materials, Hunan Key Laboratory of Mechanism and Technology of Quantum Information, College of Aerospace Science and Engineering, National University of Defense Technology, Changsha 410000, China

[2]School of Computer, State Key Laboratory of Complex & Critical Software Environment, National University of Defense Technology, Changsha 410000, China

[3]College of Electronic and Information Engineering, Tongji University, Shanghai 201804, China

[4]School of Material Science and Engineering, Peking University, Beijing 100871, China

[5]School of Physics, Central South University, Changsha 410083, China

[6]College of Science, National University of Defense Technology, Changsha 410000, China

[7]School of Advanced Materials, Peking University, Shenzhen Graduate School, Shenzhen, Guangdong 518055, China

[8]These authors contributed equally.

\*Corresponding authors: K.X.(xukelele@nudt.edu.cn); S.W.(wangshanshan08@nudt.edu.cn)

## Abstract

The emergence of deep learning (DL) has provided great opportunities for the high-throughput analysis of atomic-resolution micrographs. However, the DL models trained by image patches in fixed size generally lack efficiency and flexibility when processing micrographs containing diversified atomic configurations. Herein, inspired by the similarity between the atomic structures and graphs, we describe a few-shot learning framework based on an equivariant graph neural network (EGNN) to analyze a library of atomic structures (e.g., vacancies, phases, grain boundaries, doping, etc.), showing significantly promoted robustness and three orders of magnitude reduced computing parameters compared to the image-driven DL models, which is especially evident for those aggregated vacancy lines with flexible lattice distortion. Besides, the intuitiveness of graphs enables quantitative and straightforward extraction of the atomic-scale structural features in batches, thus statistically unveiling the self-assembly dynamics of vacancy lines under electron beam irradiation. A versatile model toolkit is established by integrating EGNN sub-models for single structure recognition to process images involving varied configurations in the form of a task chain, leading to the discovery of novel doping configurations with superior electrocatalytic properties for hydrogen evolution reactions. This work provides a powerful tool to explore structure diversity in a fast, accurate, and intelligent manner.

**Keywords:** Graph neural network, Transmission electron microscopy, Defects, Atomic structures

## 1. Introduction

Revealing the material structure at atomic resolution is the cornerstone to establishing structure-property correlation for advanced material design [1-4]. Theoretically, the material properties are determined by the atoms and their spatial arrangements [5,6], which nowadays can be directly visualized by advanced transmission electron microscopy (TEM) in picometer-level precision [7,8]. Assisted by the TEM technique, diverse atomic structures like topological defects [9,10], point defects [11-13], grain boundaries [14-16], and stacking configurations [17-19] have been extensively investigated, promoting the discovery of peculiar physical properties and their potential applications in electronics, optics, and energy storage, etc.[20-22] Essentially, an atomic-resolution TEM image is the 2D projection of 3D structure [7,23], where not only are the top-view atom arrangements visualized but the composition information of each atomic column can also be reflected in some specific imaging modes, such as annular dark-field scanning transmission electron microscopy (ADF-STEM). Therefore, once a TEM image is obtained, the analyzing process by a domain expert involves the following steps: (i) pinpointing each projection point; (ii) determining the coordination environment based on the spatial distribution of the atomic columns; and (iii) identifying the microstructure with domain knowledge after evaluating the local gray values of each atomic column and their spatial arrangements (Fig. 1a). However, the detection, classification, and quantitative analysis of all the atomic structures by humans requires immense labor with biased judgment, which is particularly pronounced for high-throughput TEM investigation. Therefore, an automated method with promoted accuracy, efficiency, and statistical rigor is highly expected.

Fortunately, DL algorithms have gradually displayed superiority in the image processing fields [24-26], owing to the exceptional ability to map the relation between images and contents. Progress has been achieved in vacancy identification [27-29], phase segmentation [30-32], and crystal symmetry determination [33], where even massive defect recognition and statistical analysis of a low signal-to-noise ratio dataset can be realized [34,35]. Typically, the DL algorithms can be divided into unsupervised and supervised strategies depending on whether annotated training datasets are required. Unsupervised learning supports the discovery of inherent relations among unlabeled data using clustering or association without upfront human intervention, which has been employed in the classification of relatively ordered structure motifs like vacancies and doping [29,36,37]. However, it may encounter difficulties when dealing with microscopy datasets containing complex structures due to a lack of predefined target variables. In contrast, supervised learning can achieve superior prediction accuracy and flexibility over a wider range of complex tasks when the model is trained with sufficient manually labeled data or simulation data with ground truth [38-40]. However, applying TEM simulation images with labels generated by software as the training dataset is only suitable for the construction of simple defects (e.g., discrete vacancies and doping) but is difficult to batch produce complex and diversified atomic structures (e.g., grain boundaries and aggregated vacancies with flexible lattice deformation), while the use of experimental TEM images as the training sets faces the limitations of high data acquisition cost and time-consuming manual annotation [41,42]. Furthermore, a supervised learning model that is trained by fixed-size patches generally lacks compatibility and efficiency when local regions display a range of structural variations, leading to overfitting issues and the consumption of computing sources [43]. Therefore, it is highly desirable to develop a versatile supervised DL framework with an effective and flexible data representation approach that enables to swiftly and accurately identify various atomic structures using a limited training dataset for practical TEM image analysis.

Given the similarity between crystal structures and graphs, atomic-resolution TEM images can be effectively represented by a set of nodes and edges, (Fig. 1b), where nodes are atoms, and edges describe the coordination between atoms via an adjacent matrix. During graph learning, each node aggregates messages and updates its embedding from the neighboring nodes in vector representations at every convolution layer. Iterative message passing enables each node to build representations that capture larger and more complex patterns for the classification of various atomic structures at node or graph level [44-46]. Compared to those image-driven DL models, graph learning offers three distinctive advantages in processing atomic-resolution TEM images. Firstly, graph representation that only contains node and connection information greatly simplifies the data structure, guiding the neural network to concentrate on atoms and their coordination manners, which are closely related to the atomic-scale structural identification, rather than wasting attention on those pixels representing the vacuum, thus outputting good inference results with less training data, time and computing power consumption. Secondly, the receptive field in graph learning can perfectly adapt to the structural variation according to connections between nodes, thus promoting flexibility for feature extraction against structural diversity. Finally, graph representation shows good interpretability as the model outputs well align with the atomic structure, facilitating a direct knowledge acquisition from the experimental data.

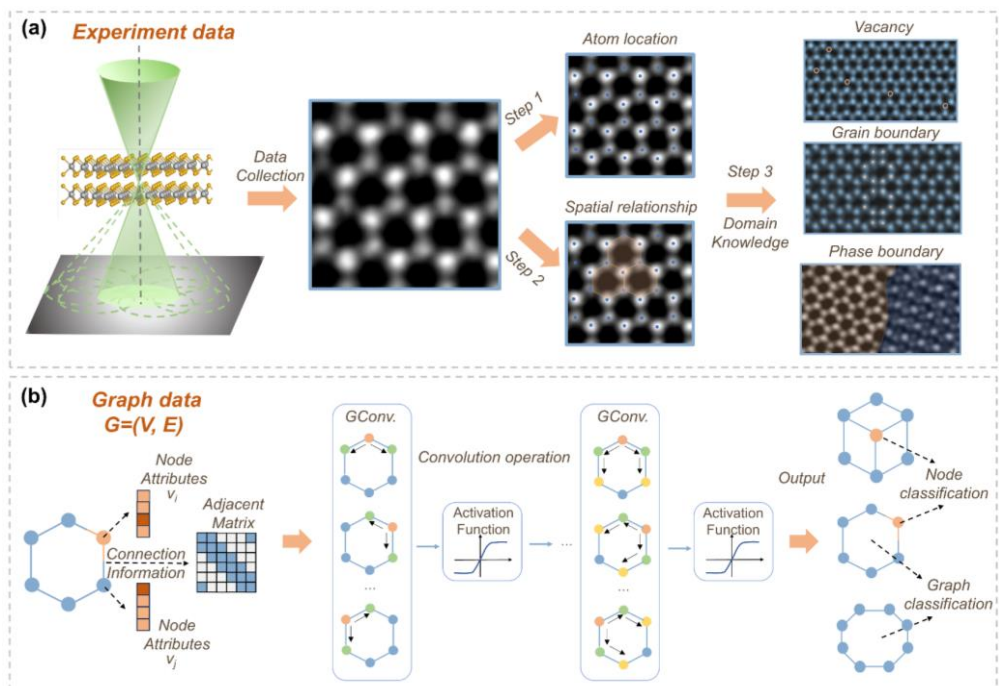

**Fig. 1.** Schematics showing the similarity between the manual analysis process of an ADF-STEM image and the graph neural network (GNN) modeling process. (a) Manual analysis process by taking an ADF-STEM image of $MoS_2$ as a typical example. The blue dots represent the atomic centroid after pinpointing the atom location (Step 1) and the orange semitransparent area shows the hexagonal structure of $MoS_2$ after determining the spatial relationship (Step 2). Different atomic structures like vacancies (orange circles), grain boundaries (orange dots), and stacking configurations (orange and blue semitransparent areas) therefore can be identified after evaluating the local gray values of each atomic column and their spatial arrangements (Step 3). (b) Graph structure and the GNN modeling process. Graph data is defined as a set of nodes (V) and edges (E), where the relation between nodes is described via an adjacent matrix. GConv. refers to the graph convolution process and the black arrows represent the message-passing directions, where each graph convolution layer enables the nodes to aggregate messages and update its embedding from their nearest neighbors (from the orange node to the green node after one layer of convolution, and to the yellow node after an additional layer of convolution). After a series of graph convolution operations, different nodes or graphs can be classified according to the output feature vectors mapped by the activation function.

Herein, by further incorporating the geometric rotation invariance in the Euclidean plane, we build a graph learning framework based on an EGNN architecture for atomic-resolution (S)TEM investigation. With few-shot training, our framework demonstrates excellent flexibility and robustness in identifying a library of atomic structures (including vacancies, stacking configurations, doping, etc.), particularly excels in the case with flexible lattice distortion (like the aggregated vacancy lines) compared to those image-driven DL models. In addition, the intuitiveness of graphs helps quantitative and batch extraction of atomic structure features (e.g., defect concentration, local coordination manner, atomic-scale mixing state, etc.), thus statistically revealing a homophilic and anisotropic self-assembly behavior of vacancy aggregates in semiconducting $MoS_2$ monolayers driven by electron beam irradiation. Furthermore, we develop a versatile model toolkit by assembling the trained EGNN sub-models for single atomic structure identification, which enables to unclose comprehensive configurations in atomic-resolution images in the form of a task chain. It uncloses a novel Pt doped on defective grain boundary structure with superior catalytic properties for hydrogen evolution reactions, demonstrating the machine learning contribution to the new structure and knowledge discovery in material science.

## 2. Results and discussion

The core concept of our framework is rooted in graph learning, which is illustrated by taking the recognition process of pristine Mo, 2S (double S atoms overlapped vertically), and the defective S vacancy sites (one S missing at the 2S site) in an ADF-STEM image of monolayer $MoS_2$ as a study case (Supplementary Fig. S1). To transfer raw image into graph data, in the first step, an optimized U-Net accompanied by Otsu and circular Hough transform algorithm (Fig. 2a) was employed to pinpoint the atomic columns for feature point map generation (Fig. 2b). The second step is planar graph generation, where we programmatically connected the identified atomic columns with their neighbors via Delaunay triangulation in a tuned edge length (Fig. 2c). In the third step, graph samples were automatically extracted by searching the first or second nearest neighbors of each node from the generated planar graph, where all the nodes in the extracted graph samples were then assigned a spatial coordinate and local gray value as their initial embeddings (Fig. 2d). Finally, a trained GNN was employed to aggregate the feature vector of each node from its neighbors and update the preset one into a high-dimensional representation (Fig. 2e). After multi-layer aggregation, the high-level feature vector of each graph sample was then input into the Softmax classifier to obtain the corresponding probability for classification at node or graph level (Fig. 2f).

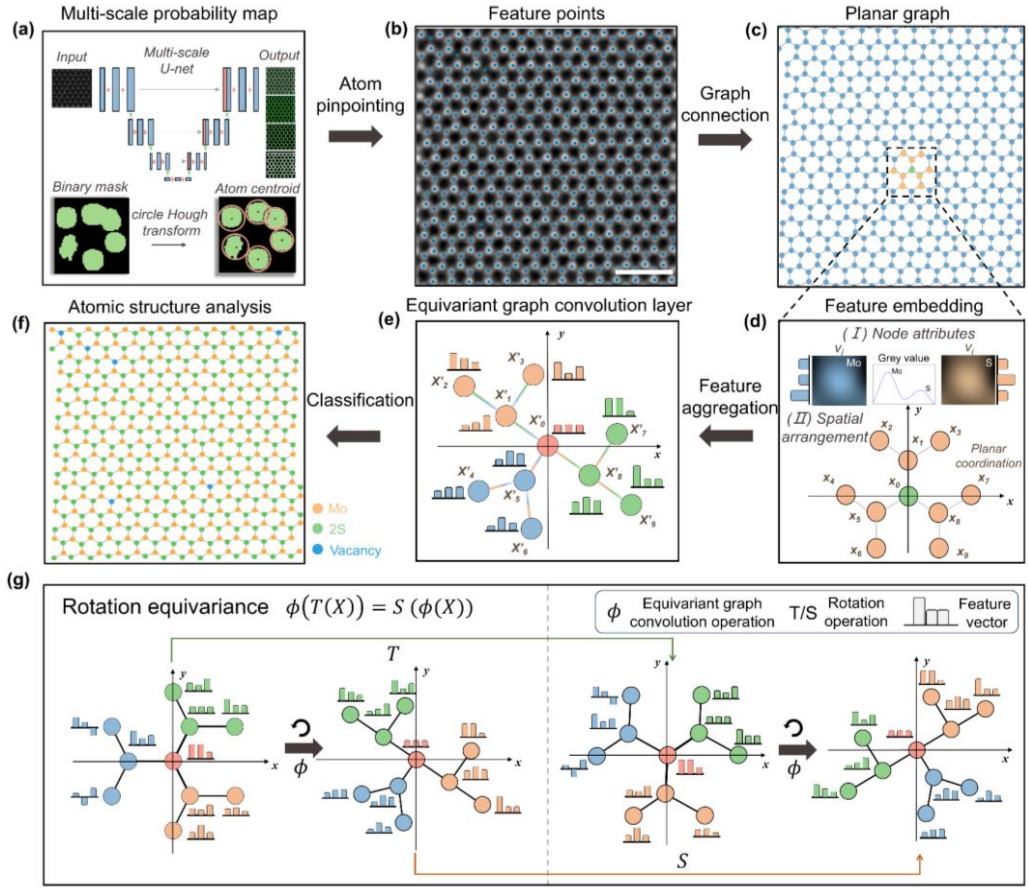

**Fig. 2.** DL workflow to identify atomic structures by taking an ADF-STEM image of monolayer $MoS_2$ containing vacancies as a typical example. (a) Illustration of the atom pinpointing architecture. The top half shows a multi-scale U-Net analysis process that generates predictions from multiple resolutions. The bottom half shows the visual results of the output probability map from the multi-scale U-Net after experiencing binarization via the Otsu algorithm and centroid detection via the circular Hough transform algorithm sequentially. (b) ADF-STEM image of monolayer $MoS_2$ with pinpointed feature points. Scale bar: 1 nm. (c) Programmatic graph generation via Delaunay triangulation connection. (d) Graph samples extracted by searching the first- and second-nearest neighbors with any detected node as the graph center (green dot) for initial feature embedding. The gray value of each atomic column is embedded in its corresponding node in the form of a feature vector. The planar coordination of each node is extracted by taking the graph center as the origin. (e) Schematic exhibiting the output graph structure after feature aggregation via the trained EGNN, where the initial feature embedding of each node has been updated by aggregating their neighbors' features. (f) Identification results corresponding to the atomic sites of Mo, 2S, and S vacancies in the form of the planar graph from (b). (g) Schematic showing the rotation equivariance of EGNN, which can be described as an operation where the output graph feature embedding (the second panel) synchronously changes with the input graph feature embedding (the first panel) after experiencing rotation operation ("T" corresponds to the rotation operation of the input graph embedding and "S" corresponds to the rotation operation of the output graph embedding).

In the first step of our workflow, a multi-scale U-Net [47] was applied to simultaneously extract features at multiple resolutions through downsampling the intermediate feature maps of the input ADF-STEM image, providing explicit guidance from distinct prediction branches to enhance the performance of the vanilla model (Supplementary Fig. S2). The as-generated probability map was subsequently binarized as foreground and background regions by an Otsu algorithm [48], which can adaptively optimize the segmentation threshold by maximizing the between-class variance (Supplementary Fig. S3) in terms of feature distribution variation between the probability maps. Finally, a circular Hough transform algorithm [49] was employed to transform the as-detected foreground into a circular object, and the circle center was thereafter identified as the centroid of the atomic columns. The optimized algorithm enables pinpointing atomic columns with an F1 score of 99.8%, even for the images suffering from astigmatism or surface contamination (Supplementary Fig. S4), laying a good foundation for the subsequent graph construction.

A general connection method based on Delaunay triangulation was then implanted to locate the neighbors of each feature point and create edges for planar graph generation, with which the graph samples can be directly extracted via local neighborhood search in a universal setting. Note that, to avoid disconnection or overcounting the distant neighbors during the edge generation process, we set a length threshold for the Delaunay edge according to the lattice parameters, where any edges that exceed this length were removed, ensuring the planar graph to be maximumly consistent with the crystal structure (Fig.2c and Supplementary Fig. S5). Once graph samples have been extracted, we assigned every node a flattened 9-dimensional feature based on the local gray value of the corresponding atomic column, which contains the composition information (Fig. 2d and Supplementary Fig. S6), and a planar coordinate to describe its spatial arrangements.

Since rotation transformation will not break the geometry symmetry but change the graph representation in the vector space, the graph feature learning may encounter difficulties when presented with structural variations significantly different from the training data. Therefore, a geometric equivariance with respect to the rotation transformations in the Euclidean plane is highly desired to enhance the robustness of the as-learned graph features. Given that, in the final step of our framework, we employed an emerging model named EGNN [50] to extract graph features under the equivariant restriction against rotation variation. Different from vanilla GNN models, the graph convolution layer in our EGNN model is designed to aggregate the spatial position information of each node during the edge operation process, enabling a simultaneous feature update of embedded values and spatial position against rotation transformation to make the output graph representation more robust (Fig. 2g and Supplementary Fig. S7).

This framework readily extends to identify boundaries and interfaces across various materials involving different compositions, dimensionalities, lattice symmetries, and functionalities (Supplementary Fig. S8 and Fig. S9). Examples include *2H-3R* stacking boundary and grain boundary identification of atomically thin $MoS_2$ (Fig.3a (i), (ii)), phase boundary segmentation (between O3 and rock salt phase) in a layered cathode lithium-ion battery material of NCM 811 (referring to $LiNi_{0.8}Co_{0.1}Mn_{0.1}O_2$) (Fig.3a (iii)), and hetero-interface recognition between bulk elemental materials of aluminum (Al) and silicon (Si) (Fig.3a (iv)). We take the ADF-STEM image segmentation of 2H, 3R, and 1H stacking configurations in $MoS_2$ as a typical study case (Fig. 3b). By searching the first nearest neighbors of each identified node from the generated planar graph (Supplementary Fig. S10), graph samples corresponding to different stacking configurations (Fig. 3c) can be automatically extracted despite the structure diversity, which align well with their crystal structures, indicating the flexibility and intuitiveness of Delaunay triangulation for the graph generation. With few-shot training (each configuration with ~200 graph samples extracted from two pieces of ADF-STEM image, Supplementary Fig. S11), the EGNN model attains a steady decrease in loss (obtained by the cross entropy loss function, see details in Methods) and shows a minor mean squared error (MSE) of 0.019 between the training and validation loss values (Fig. 3d), proving the validity of EGNN model for few-shot learning in graphs. Although the loss obtained by the CNN model (ResNet 18) [51] also exhibits a rapid convergence trend after training several epochs, a large gap between the validation and training loss values is observed with an MSE of 0.341, indicating a pronounced overfitting issue. It may be ascribed to the excessive redundant information extracted from image patches in such a limited training dataset. As a result, the EGNN model achieves an impressive F1 score of over 0.98 for the classification of stacking configurations in this test case (Supplementary Fig. S13a), which outperforms the CNN model showing an F1 score of 0.60 (Supplementary Fig. S12). More importantly, nearly 8000 times reduced computing parameters are performed via graph learning, displaying 3.9 times promoted inference efficiency compared to that using the image-driven model (Fig. 3e). Fig. 3f shows the end-to-end visualization results obtained by our framework, whereby a clear boundary can be identified between different stacking configurations (Supplementary Fig. S13b). Note that, all the misidentifications are primarily situated at the stacking boundary, which are inherently indistinguishable even for a domain expert.

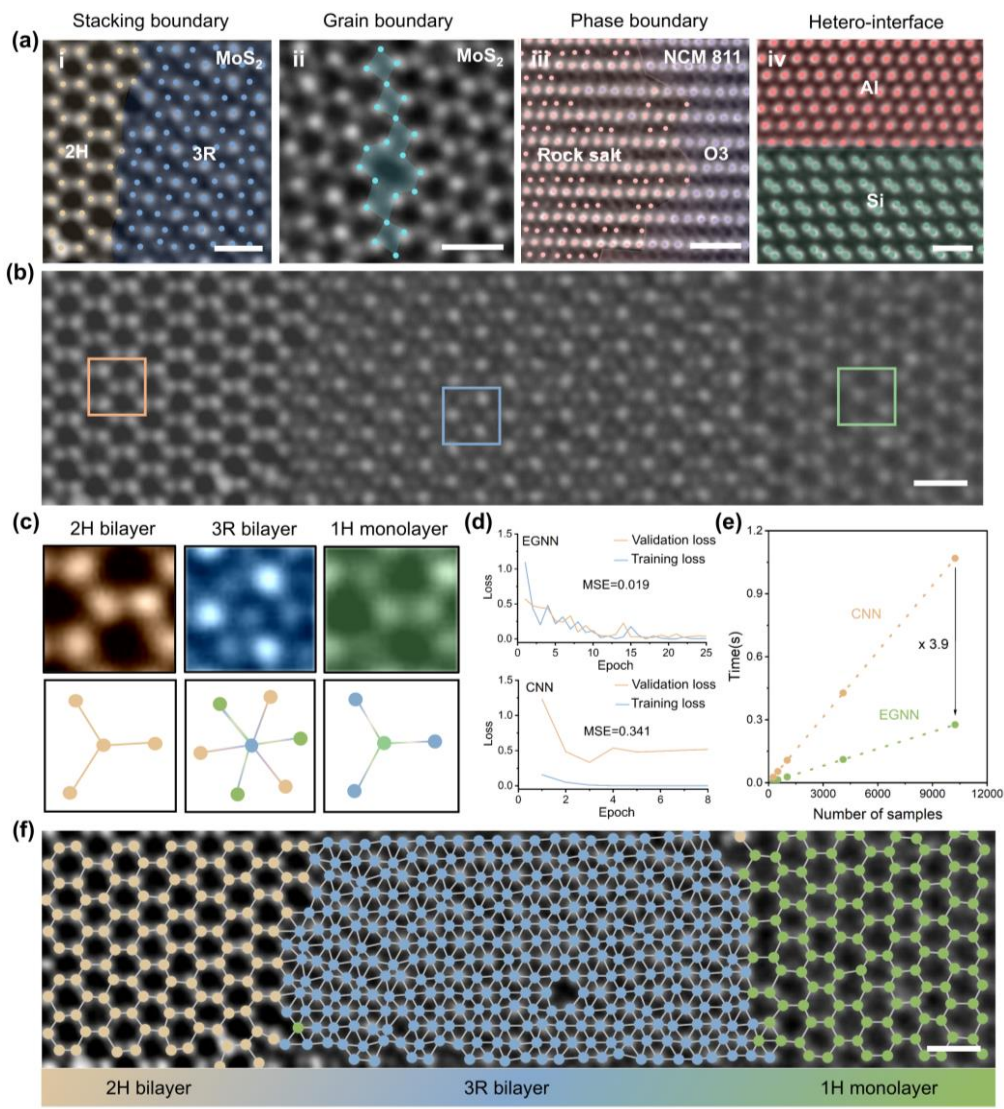

**Fig. 3.** Identification of boundaries and interfaces across various material systems. (a) Visualization results of the *2H-3R* stacking segmentation in bilayer MoS$_2$ (i), grain boundary extraction in monolayer MoS$_2$ (ii), phase boundary segmentation of NCM811 (iii), and interface identification of Al/Si heterostructure (iv), respectively, inferred by our DL algorithm. The orange, blue, cyan, pink, purple, red, and green dots represent the identified atomic columns corresponding to the *2H* stacking, *3R* stacking, grain boundary, rock salt phase, O$_3$ phase, Al, and Si. Scale bars: 0.5 nm. (b) ADF-STEM image of a bilayer-monolayer MoS$_2$ lateral heterostructure, which involves 2H (bilayer), 3R (bilayer), and 1H (monolayer) stacking configurations from left to right. Scale bar: 0.5 nm. (c) Image patches of different stacking configurations from (b). The orange, blue, and green nodes represent the Mo+2S, 2S, and Mo atomic sites, respectively. (d) Training and validation loss generated by EGNN and CNN models. (e) Plot showing the inference time that varies with the image patch or graph sample numbers using CNN and EGNN models. (f) Visualization result from (b) by our framework. Scale bar: 0.5 nm.

In addition to the relatively ordered microstructures studied above, our framework verifies superiority in analyzing defective structures with prominent and flexible lattice distortion. A typical example is the recognition of aggregated S vacancy lines in monolayer $MoS_2$, which are in-situ produced by the electron beam irradiation during the high-resolution TEM (HR-TEM) imaging process, whose structures would evolve and diversify as the irradiation time prolongs (Supplementary Fig. S14). Distinct from the discrete S vacancies, the aggregated S vacancy lines exhibit significant geometry variations in both length and width, inducing flexible and diversified lattice distortion as large as 20.8% along the armchair direction (Supplementary Fig. S15), which poses challenges for conventional computer vision algorithms to effectively extract features from convolution kernel or other representations in fixed patches (Supplementary Fig. S16). On the contrary, graph representation can fix this limitation as the receptive field can well adapt to the structural variation once a connection between each node has been set. More importantly, the cognitive ability of the EGNN model in the spatial position of each node enables to detect the aggregation status of vacancies, which well aligns with the human cognitive criteria and endows the algorithm with interpretability and intuitiveness.

Likewise, we trained an EGNN model using only one HR-TEM image of the 0 s irradiation sample (image size: 2048×2048) and tested it on the 10 s one (Fig. 4a), which has already experienced severe lattice disorder and geometry variation (Fig. 4b). By searching the second-order neighbors, graph samples corresponding to different defect configurations (typical examples shown in Fig. 4d) can be flexibly extracted in the as-generated planar graph (Fig. 4c) regardless of the image patch size. Therefore, graph samples in varied geometry can be well separated into three clusters with a distinctive boundary in the feature space (Fig. 4e), showing the highest F1 scores of 97.6% and 97.5% for the classification of aggregated line and discrete point defects, respectively (Fig. 4f). Notably, removing the spatial coordinates of each node will eliminate the model's ability to perceive the aggregate state of the vacancies, thus resulting in an increased proportion of misidentifying discrete point vacancies as the linearly aggregated ones (Fig. 4f and Supplementary Fig. S17a). Other state-of-the-art architectures like ResNet 18 and Zernike polynomial (ZP) [37] also moderately underperform in this task, where the F1 scores for line and point defect recognition using these two models decrease to 92.5%, 84.7%, and 85.7%, 51.1%, respectively (Fig. 4f and Supplementary Fig. S17b and c).

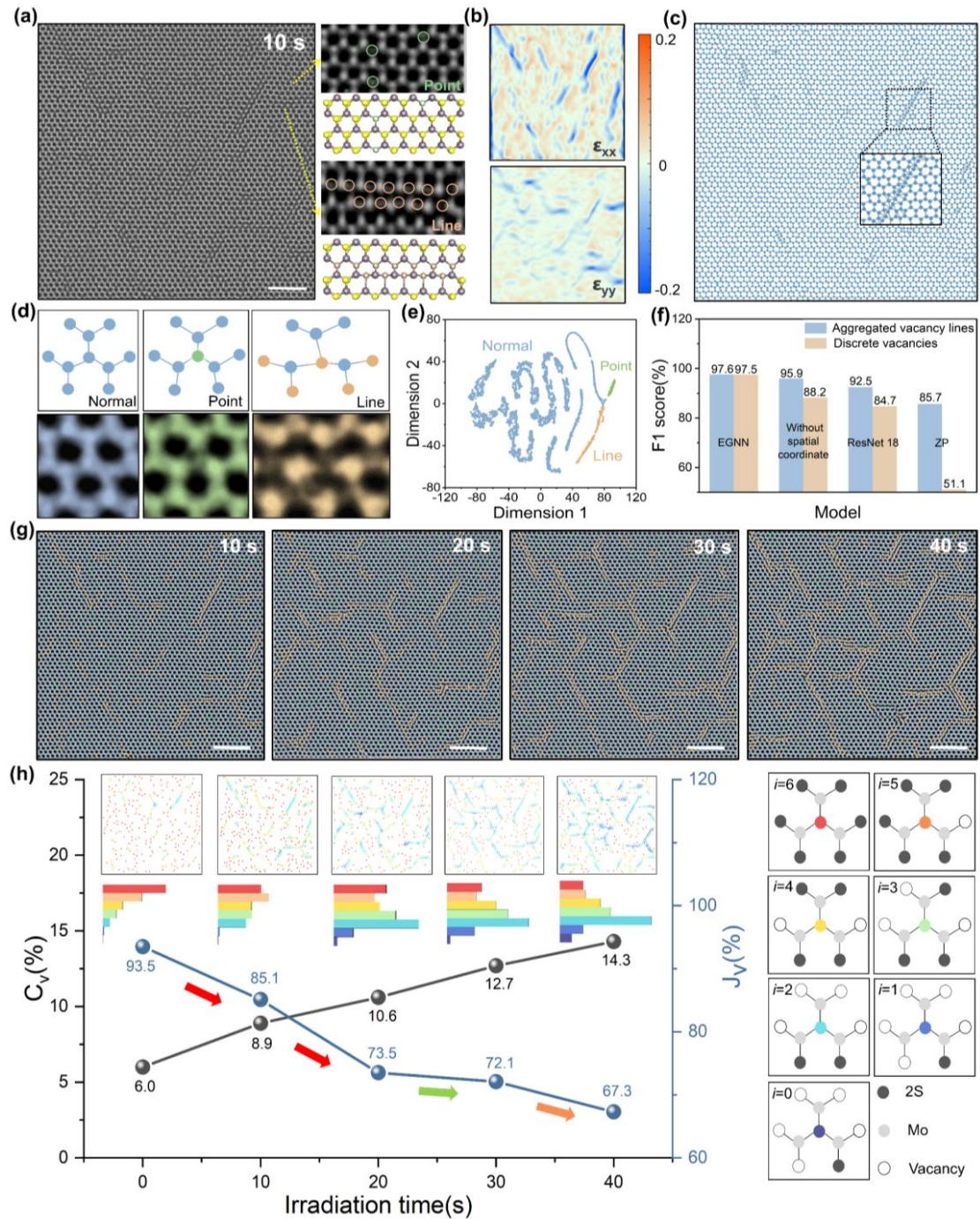

**Fig. 4.** Automatic analysis of aggregated S vacancy lines with flexible and varied lattice distortion. (a) HR-TEM image of monolayer $MoS_2$ after 10 s e-beam irradiation. The right panels present the enlarged views of the discrete S vacancies and aggregated S vacancy lines with their corresponding atomic models from top views. Scale bar: 2 nm. (b) Strain map of (a) in the horizontal and vertical directions through geometric phase analysis. The color scale ranges from -0.2 to 0.2. Obvious compression strain (blue area) can be observed around the vacancy lines. (c) Planar graph representation generated from (a). (d) Typical graph samples centered on different types of atomic columns with their corresponding image patches from (a) to illustrate the flexibility of graph representation. The blue, green, and orange dots represent the normal atomic columns, discrete vacancies, and linearly aggregated vacancies, respectively. (e) T-SNE outputs in feature space generated by the trained EGNN model from (a). (f) Performance metrics of EGNN model and other representative algorithms tested on (a) for comparison. (g) End-to-end recognition results by our framework tested on the HR-TEM image series of monolayer $MoS_2$ after experiencing 10 s, 20 s, 30 s, and 40 s e-beam irradiation. Scale bars: 2 nm. (h) Plots showing the vacancy concentration ($C_v$) and the alloying degree of S vacancies ($J_v$) as a function of irradiation time. The inset images show the quantity variation of vacancies that are paired with 6 (red), 5 (orange), 4 (yellow), 3 (green), 2 (cyan), 1 (blue), and 0 (dark blue) 2S atomic columns as the irradiation time prolongs, which can be directly extracted from the graph outputs by counting the number of disulfide atomic columns on the second nearest neighboring sites of each S vacancy (illustrated in the right panels, $i$ is the coordination number ranging from 0-6). The inset bar charts show the corresponding ratio variation of the vacancies paired with different numbers of 2S atomic columns. The red, green, and orange arrows illustrate the rapid decline stage, the platform stage, and the moderate decline stage, respectively.

To further manifest the robustness of our frameworks, a series of in-situ HR-TEM images suffered from varied lattice distortion was employed for end-to-end evaluation. The recognition ability of our framework remains intact regardless of the increased lattice distortion (Fig. 4g and Supplementary Fig. S18), realizing F1 scores of 98.3% and 87.3% for the recognition of linearly aggregated and discrete vacancies in the 40 s irradiation sample when the proportion of aggregated vacancies (ratio between aggregated vacancies and the total S atom number) has already increased from 3.5% (corresponding to the training data) to 14.3%. However, multi-scale U-Net alone fails to work in this situation (Supplementary Fig. S19), and so do the other classic end-to-end DL models like YOLO based on object detection [52] (Supplementary Fig. S20). Note that, even though the number of graph samples in this task increases to more than 5000, end-to-end computing of one image by our framework only takes 0.56 s, showing a 5.5 times improvement in efficiency compared to that by CNN (Supplementary Table S2), realizing the real-time TEM investigation for dynamic structural evolution (Supplementary Movie S1).

Due to the inherent intuitiveness of graphs, our framework can directly and quantitatively extract structure-related parameters (e.g., defect concentration, local coordination number, atomic-scale mixing state, etc.) from the HR-TEM image series in Fig. 4g involving ~27,000 atoms, thus statistically revealing self-assembly behavior of S vacancy aggregates in monolayer $MoS_2$ driven by electron beam irradiation, which may help precise defect engineering and electronic band modulation at the atomic level for advanced nanodevice design [53]. Generally, the vacancy concentration shows a linear increase with the irradiation time from 6% in the initial state to nearly 15% after undergoing 40 s irradiation (black line in Fig. 4h). A statistical analysis of the coordination number of S vacancies was then achieved by quantifying the number of 2S atomic columns located at the second nearest neighboring sites of each S vacancy (denoted as $i$, ranging from 0 to 6, right schematics in Fig. 4h), showing a decrease of the dominant $i$ from 6 (red columns) to 2 (cyan columns) as the S vacancy proliferate (inset bar charts of Fig. 4h). It indicates vacancy aggregation and an increasingly significant deviation of the local stoichiometric ratio of Mo/S from 1:2. We further investigate the irradiation time-dependent mixing state evolution between S vacancies and the pristine 2S sites at the atomic scale by introducing a parameter called 'alloying degree' ($J_v$) [54]. All information needed for $J_v$ calculation can be obtained from the model outputs (Supplementary Fig. S21). Two interesting phenomena were unclosed (blue line in Fig. 4h): (i) $J_v$ is always less than 100% and continuously decreases with the irradiation time, indicating in increasingly pronounced homophilic assembly manner of S vacancies (i.e., vacancies prefer to aggregate with vacancies rather than 2S sites). (ii) The decline rate of $J_v$ displays three stages, indicating the self-assembly process of vacancy lines from nucleation to anisotropic growth (Supplementary Fig. S21). The rapid decline stage (red arrows) corresponds to the aggregation of isolated vacancies to form short and narrow baby vacancy lines. Then a platform stage (green arrow) follows corresponding to the extension of vacancy lines along the zigzag lattice directions in MoS$_2$. Finally, a moderate $J_v$ decline stage (orange arrow) comes corresponding to the broadening of the as-grown long and narrow vacancy lines along the the armchair lattice directions.

In practical situations, materials encompass not just a single type of defect but a library of defective configurations [55]. However, solely relying on a specific model fails a comprehensive analysis of complex systems, and developing a new model is not always feasible due to the scarcity of training data. Herein, we provide a model toolkit by assembling the trained sub-models to support an all-around exploration of diverse atomic configurations in the form of a task chain. Proof of this concept is exemplified by analyzing an ADF-STEM image of Pt-doped defective MoS$_2$ that coexists with a range of atomic configurations such as doping, grain boundaries, stacking configurations, and vacancies (Fig. 5a), where there is generally no training dataset available for such a specific level of structural complexity. To screen out all the atomic configurations, in the first step, we additionally trained a doping configuration corresponded EGNN sub-model and assembled it with the previously trained ones (including the EGNN models for identifying vacancies, grain boundaries, and stacking configurations, etc.) to establish a versatile model toolkit. Secondly, we transferred the input ADF-STEM image into a graph representation for feature extraction. Finally, a task chain was built by combining EGNN sub-models corresponding to identifying doping, stacking configurations, grain boundaries, and vacancies in sequence, with which the diversified atomic structures can be comprehensively analyzed by overlaying the recognition results from different models.

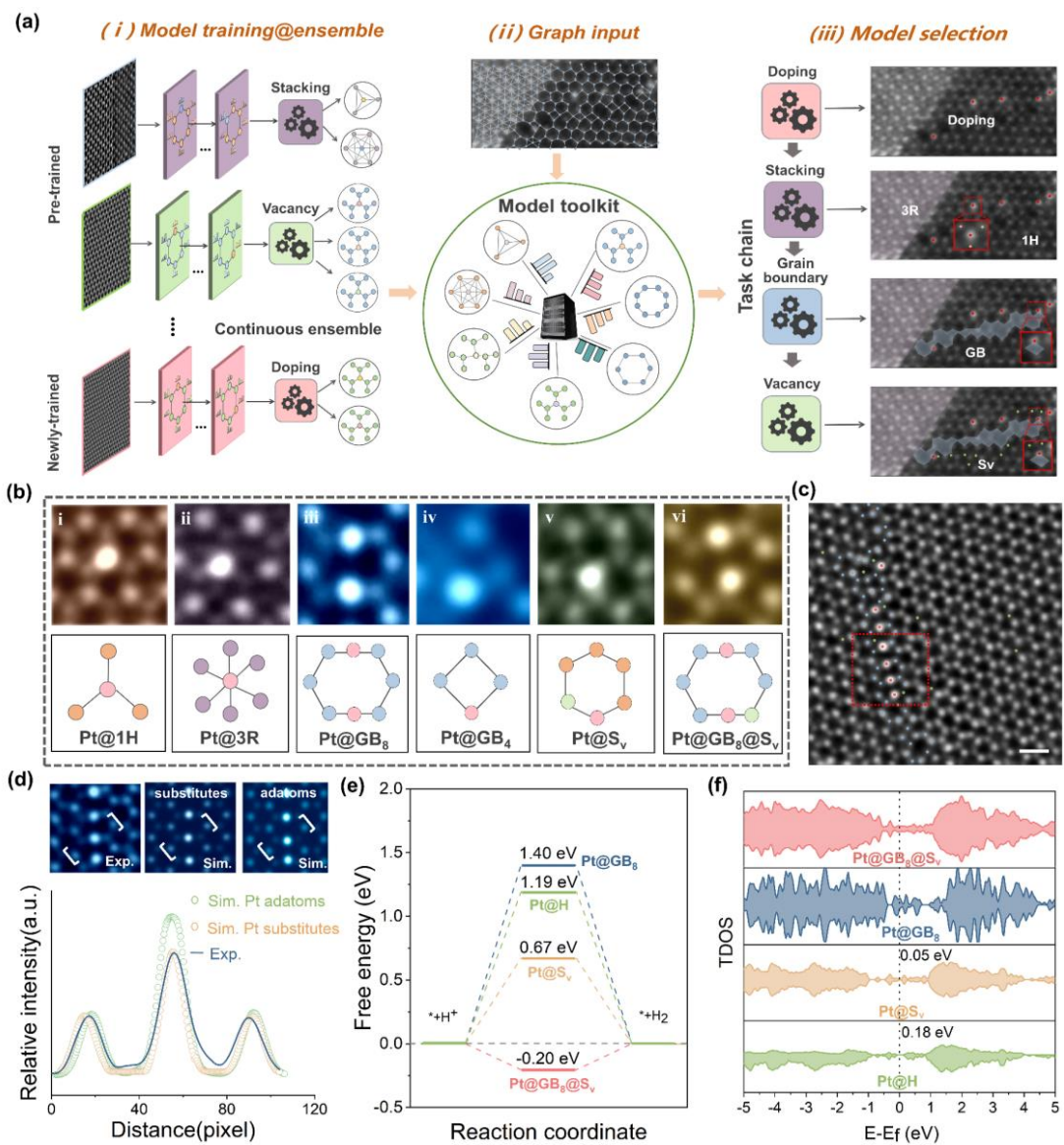

**Fig. 5.** Comprehensively exploring diversified atomic structures by integrating different sub-models in the form of a task chain. (a) Schematic showing the sequential analysis process of various types of atomic configurations in the form of a task chain by employing different EGNN sub-models. An ADF-STEM image of Pt-doped MoS$_2$ containing a range of atomic configurations such as doping, grain boundaries, stacking configurations, and vacancies is exemplified here. The red squares in the right panels show the enlarged views of typical Pt doping configurations after stepwise overlaying the recognition results from different models' outputs. (b) Extracted patches of different Pt doping configurations with their corresponding graphs. The pink, orange, purple, blue, and green nodes represent Pt, 1H monolayer, 3R stacking, grain boundaries, and vacancies, respectively. (c) ADF-STEM image showing the Pt atom situated at the 8-fold rings of Mo sites in the form of an aggregated row and its corresponding identification results. The red, blue, and green dots represent the doping, grain boundaries, and vacancies. (d) Intensity line profiles measured between the experiment and simulation images. The experimental image is extracted from the red square marked in (d). We used two simulation models, Pt substitutes of Mo and Pt adatoms on Mo around the 8-fold rings of grain boundary, to compare their intensity line profiles

with the experimental data, where the intensity line profile corresponding to the Pt substitutes of Mo fits better with the experiment data. (e) Calculated hydrogen adsorption free energy for different Pt doping configurations. (f) TDOS plots calculated from different Pt doping configurations.

This process can be extended to batch analyze a series of data involving $10^4$ atoms, which is expected to discover novel configurations from the models' outputs by investigating the combinations among the 'known' ones (e.g., if different EGNN sub-models recognize the same location in a micrograph for the existence of corresponding microstructures, it indicates the potential emergence of a new 'unknown' configuration here that is constructed by the combination of these submodel-corresponded 'known' single structures). A detailed investigation of the Pt doping sites and their local atomic environments was realized by stepwise overlaying and analyzing the identification results of Pt doping with that of the others (stacking, grain boundary, and vacancy corresponded models, see details in Supplementary Fig. S22). Beyond the extensively investigated doping configurations [56] like the Pt located at the pristine 1H monolayer (Pt@1H) and 3R-stacked bilayer $MoS_2$ (Pt@3R) with no defect around (Fig.5b (i) and (ii)), our framework discovered a larger range of doping diversity: Pt located at the 8-fold rings of grain boundaries (Pt@$GB_8$, Fig.5b (iii)), Pt located at the 4-fold rings of grain boundaries (Pt@$GB_4$, Fig.5b (iv)), Pt located at the basal plane of 1H monolayer and at 8-fold rings of grain boundaries with S vacancy around (Pt@$S_v$ in Fig.5b (v), Pt@$GB_8$@$S_v$ in Fig. 5b (vi)). In addition, when tracing the distribution of different Pt@GB configurations (Supplementary Fig. S22), it can be found that Pt atoms are prone to locate at the 8-fold rings of grain boundaries and an aggregated Pt row along the grain boundary with vacancy around are detected (Fig. 5c). By further analyzing the intensity line profiles between the simulation and the experimental data, these Pt atoms are verified to be the substitutes of Mo rather than the adatoms (Fig. 5d).

Since Pt is highly regarded as a top-performing dopant for promoting the electrocatalytic performance of 2D $MoS_2$ in hydrogen evolution reaction (HER) [2], the discovery of these novel doping configurations motivates us to investigate their potential catalytic activity for hydrogen production. DFT calculations were conducted to examine the adsorption free energy of H* ($\Delta G_H$) of Pt@$GB_8$ and Pt@$GB_8$@$S_v$. Pt@$S_v$ and Pt@1H (Pt situated in the metal site) were employed for performance comparisons, where 2S or $S_v$ sites around the Pt atoms were chosen as the adsorption sites according to the previous works [57] (Supplementary Fig. S23). Fig. 5e shows that both Pt@$GB_8$@$S_v$ and Pt@$S_v$ exhibit superior $\Delta G_H$ of -0.20 eV and 0.67 eV than that of Pt@$GB_8$ (1.40 eV) and Pt@1H (1.19 eV), indicating a dominant role of the vacancy-Pt pair in promoting the catalytic properties. Note that, the vacancy-Pt pair situated in the 8-fold rings of grain boundaries pushes $\Delta G_H$ close to 0 eV, which is more beneficial to compromise the reaction barriers in the adsorption and desorption steps. Electrical conductivity is another key factor for catalytic activity. Good electrical conductivity ensures efficient electron transfer during the HER process. Fig. 5f shows the total density of states (TDOS) of these four doping configurations. Both Pt@GB$_8$ and Pt@GB$_8$@S$_v$ introduce electronic states around the Fermi level, showing a semiconductor-to-metal transition. Although the position of the valence band moves downward when forming a vacancy-Pt pair in the Pt@1H, the Pt@S$_v$ configuration still exhibits a semiconductor property with a minor band gap of 0.05 eV. The above results offer valuable insights into advanced electrocatalyst design through doping configuration modulation, showcasing our framework's contribution to new knowledge emergence.

## 3. Conclusion

To sum up, we describe a few-shot learning framework based on EGNN to investigate a series of complex atomic resolution micrographs. The focus of the graph neural network, nodes and edges, perfectly matches key information to understanding atomic-scale crystal structures, i.e., atoms and their coordination with neighbors, avoiding wasting attention on the vacuum regions (e.g., black areas in monolayer MoS$_2$ STEM pictures) that occupy a large proportion of the microscopy images. Therefore, our framework provides flexibility, robustness, and computing efficiency in identifying various atomic configurations (vacancies, phases, grain boundaries, doping, etc.) compared to image-driven DL architectures, and performs especially well for defects with flexible and varied lattice distortion, like the aggregated vacancy lines in MoS$_2$. The intuitiveness of graphs enables quantitative and straightforward extraction of structure-related parameters in batches (e.g., defect concentration, local coordination number, atomic-scale mixing state, etc.), thus revealing the evolution dynamics of vacancy lines under electron beam illumination, which has potential applications in advanced nanodevice design. We also built an extendable toolkit to explore structure diversity in the form of a task chain, contributing to the discovery of novel doping configurations with high catalytic activity. This work demonstrates a computationally efficient and versatile platform that enables automatic, accurate, fast, and batch identification of diversified atomic-scale structures with a small training dataset. It is in line with the needs of practical microscopy data analysis, helps unveil statistically grounded information, and facilitates new structure and knowledge discovery.

# 4. Methods

## 4.1 Growth and transfer of monolayer MoS$_2$

MoS$_2$ monolayers were synthesized by atmospheric pressure chemical vapor deposition (APCVD) without hydrogen introduction, according to previous reports [58]. Molybdenum trioxide (MoO$_3$, ≥99.5%, Sigma-Aldrich), sulphur (S, ≥99.5%, Sigma-Aldrich), and SiO$_2$/Si (300 nm thick SiO$_2$) chips were used as precursors and substrates, respectively. To avoid the quench of MoO$_3$ powder by S vapor during the reaction, an inner tube with MoO$_3$ powder placed inside was inserted into the outer 1-inch quartz tube, where S powder was positioned upstream. The typical heating temperatures for S, MoO$_3$, and SiO$_2$/Si substrate were ∼180, ∼300, and ∼800 °C, respectively. The growth lasted 20 minutes followed by a fast cooling. For the transfer of monolayer MoS$_2$, we firstly spin-coated a thin film of poly (methyl methacrylate) (PMMA) on the MoS$_2$/SiO$_2$/Si substrate and then floated it on a 2 mol/L potassium hydroxide (KOH) solution to peel off the PMMA/MoS$_2$ film from the SiO$_2$/Si substrate. Subsequently, the film was transferred to the deionized water three times to remove etchant residuals thoroughly. The film was scooped up by the SiN$_x$ TEM grid, naturally dried in the air, and baked on the hotplate at 180°C for 15 minutes. Finally, the PMMA scaffold was removed by submerging the TEM grid in acetone for 8 hours.

## 4.2 TEM characterization

The ADF-STEM images of MoS$_2$ were captured using an aberration-corrected JEOL ARM300CF equipped with a JEOL ETA corrector at an accelerating voltage of 60 kV. Dwell times of 5-20 μs and a pixel size of 0.006 nm px$^{-1}$ were applied. Optical conditions used a condensed lens aperture of 30 μm, a convergence semiangle of 31.5 mrad, a beam current of 44 pA, and inner−outer acquisition angles of 49.5-198 mrad. The HR-TEM images were conducted using an FEI Titan 80-300 environmental TEM under 80 kV accelerating voltage with a Gatan OneView (4k×4k) high frame rate camera.

## 4.3 Multi-scale U-Net

U-Net is a convolutional neural network architecture designed for image segmentation, which features a symmetric U-shaped structure to enable precise localization and contextual information extraction. Optimized from vanilla U-Net architecture, multi-scale U-Net is a variant of U-Net that can better capture multi-scale image features by simultaneously making predictions from different resolutions. Given an image of $p \in \mathbb{R}^{H \times W \times C}$, we use the multi-scale U-Net $\phi_{msunet}$ to extract multi-scale features $V$:

$$V = \phi_{msunet}(p) = \{v_0, v_1, v_2, v_3\} \qquad (1)$$

where $v_0 \in \mathbb{R}^{H \times W}$, $v_1 \in \mathbb{R}^{\frac{H}{2} \times \frac{W}{2}}$, $v_2 \in \mathbb{R}^{\frac{H}{4} \times \frac{W}{4}}$, $v_3 \in \mathbb{R}^{\frac{H}{8} \times \frac{W}{8}}$ correspond to the probability maps predicted at four scales of the multi-scale U-Net. To enhance the model's generalization and robustness, we employed various data augmentation techniques, including random horizontal and vertical flipping, gaussian blur, brightness, contrast adjustments, as well as random affine transformations such as translation, scaling, and rotation. During inference, we only took the probability map $v_0$ and applied the Otsu algorithm to separate the foreground and background. Then, we employed the circle Hough transform (*CHT*) for atomic center point extraction, obtaining a set of atomic center coordinates $x^0$:

$$x^0 = CHT(OTSU(v_0)) = \{x_0^0, \dots, x_{K-1}^0\}, x_k^0 \in R^2 \tag{2}$$

where $K$ denotes the number of identified atoms, and $x_k^0$ represents the two-dimensional spatial coordinates of each atom.

### 4.4 Equivariant graph neural network

EGNN extends traditional graph neural networks by incorporating equivariance properties, ensuring that the model's outputs are invariant or equivariant to the rotation transformation of the input data. For any atom as the center, we constructed the corresponding graph data $\mathcal{G} = (\mathcal{V}, \mathcal{E})$, where $\mathcal{E}, \mathcal{V}$ represents the adjacency matrix and node information, respectively. The node information $\mathcal{V}$ comprises the coordinates $x$ and features $h$, which are derived from the spatial positions corresponding to the cropped and expanded images.

For each graph sample $\mathcal{G} = (\mathcal{V}, \mathcal{E})$, the equivariant graph neural network requires three inputs, including the adjacency matrix $\mathcal{E}$, the atomic feature vector $h^l = \{h_0^l, \dots, h_{M-1}^l\}$ and the spatial coordinates of atoms $x^l = \{x_0^l, \dots, x_{M-1}^l\}$, where $M$ denotes the number of atoms in the sample, and $l \in [0, L-1]$ denotes the layer of the equilateral graph neural network. Equivariant Graph Convolutional Layer (EGCL) takes $h^l$, $x^l$, and the edge connectivity $\mathcal{E} = (e_{ij})$ as input and fuses the information to obtain $l+1$ layer features: $h^{l+1}$ and $x^{l+1}$. In short, $h^{l+1}, x^{l+1} = EGCL(h^l, x^l, \mathcal{E})$. The intra-layer operations are as follows:

$$m_{ij} = \phi_e\left(h_i^l, h_j^l, \|x_i^l - x_j^l\|^2\right) \tag{3}$$

$$x_i^{l+1} = x_i^l + \frac{1}{M-1} \sum_{j \neq i} (x_i^l - x_j^l) \phi_x(m_{ij}) \tag{4}$$

$$m_i = \sum_{j \neq i} m_{ij} \tag{5}$$

$$h_i^{l+1} = \phi_h(h_i^l, m_i) \tag{6}$$

where $\phi_e$ and $\phi_h$ are the edge and node operations respectively. $\phi_x: \mathbb{R}^{nf} \to \mathbb{R}^1$ takes the edge embedding $m_{ij}$ from the previous edge operation and outputs a scalar value. According to the different tasks, there are two classification methods: the first is classification based on the central atom:

$$y = Softmax(h_0^L), y \in \mathbb{R}^c \tag{7}$$

where $c$ represents the number of classes. Another approach is to perform classification on the graph structure:

$$y = Softmax(AGG(h^L)), y \in \mathbb{R}^c \tag{8}$$

where $AGG$ represents the aggregation operation.

**4.5 Model assembling and task chain building**

In the model database assembling scenario, let's consider having $T$ sub-models $\{\phi_1, \phi_2, \ldots, \phi_T\}$, where each model $\phi_t$ is focused on recognizing a specific atomic type or a particular defect. Given input data $D = (\mathcal{G}, \mathcal{S})$, comprising graph data $\mathcal{G}$ and a sequence of prediction tasks $\mathcal{S}$, we feed the input data $D$ into the sub-models based on the sequence $\mathcal{S}$. With the dependency relationship defined by $\mathcal{S}$ among the models, each sub-model can receive outputs from other models as its input. This facilitates information exchange between models, allowing for the subdivision of diversified structure configurations.

**4.6 Loss function**

Our framework consists of two parts: multi-scale U-Net semantic segmentation and EGNN defect classification. In the multi-scale U-Net semantic segmentation part, we used a combination of Dice loss and Cross entropy loss functions. Dice Loss was used to measure the overlap between predicted segmentation results and ground truth, particularly suitable for imbalanced categories; Cross entropy loss was used to measure the difference between model outputs and ground truth labels in classification tasks. The calculation formula is:

$$Dice\ Loss = 1 - \frac{2 \cdot TP}{2 \cdot TP + FP + FN} \tag{9}$$

$$Cross\ entropy\ Loss = -\sum_{i=1}^{N} y_i \cdot log(\hat{y}_i) \tag{10}$$

where $TP$, $FP$, and $FN$ are true positives, false positives, and false negatives, respectively. $\hat{y}_i$ is the predicted probability of the model for the $i$-th sample, and $y_i$ is the corresponding true label. In the defect classification stage of EGNN, we only used the cross-entropy loss function to optimize category prediction results.

## 4.7 Evaluation metrics

We used the F1 score, precision, and recall to evaluate the model's recognition and classification performance. Precision refers to the proportion of samples predicted as positive by the model that are actually positive. It measures the accuracy of the model in predicting positive samples. Recall refers to the proportion of actual positive samples that are correctly predicted as positive. It measures the model's ability to identify positive samples. The F1 score is the harmonic mean of precision and recall, used to balance the relationship between the two. It is particularly suitable for situations where sample categories are imbalanced. The formula is as follows:

$$Precision = \frac{TP}{TP+FP} \tag{11}$$

$$Recall = \frac{TP}{TP+FN} \tag{12}$$

$$F1 = 2 \cdot \frac{Precision \cdot Recall}{Precision + Recall} \tag{13}$$

Where TP (True Positives) represents the number of samples correctly predicted as positive class, FP (False Positives) represents the number of samples incorrectly predicted as positive class, and FN (False Negatives) represents the actual positive class samples incorrectly predicted as negative class.

## 4.8 Hyperparameters.

During the training process, our code was implemented based on the PyTorch Lightning framework. We utilized the AdamW optimizer for parameter optimization with a batch size of 32. The learning rate scheduler chosen was ReduceLROnPlateau to adapt to changes in the model's performance on the validation set. Additionally, to prevent overfitting and save training time, we introduced an Early Stopping strategy, which terminates training prematurely if the performance metrics on the validation set do not improve for consecutive 2 epochs. The entire training process was conducted for a maximum of 1000 epochs to ensure the model has sufficient opportunities for convergence and optimization.

## 4.9 DFT calculations

Vienna Ab initio Simulation Package (VASP) was employed to perform all density functional theory (DFT) calculations within the generalized gradient approximation (GGA) using the Perdew-Burke-Ernzerhof (PBE) formulation. The Gibbs free energy for each elementary step was calculated as: $G = E_{elec} + E_{ZPE} - TS$, where $E_{ZPE}$ is the zero-point energy term, and $T$ is the absolute temperature (here 298 K), $S$ is the entropy. The cutoff energy of the plane wave set as 450 eV. The convergence criterion for total energy was set to $10^{-6}$ eV and atoms were relaxed until the Hellman-Feynman forces were less than 0.001 eV Å$^{-1}$.

The 15 Å vacuum layer was added to the surface to eliminate the artificial interactions between periodic images.

## Acknowledgment

S.W. acknowledges support from the National Natural Science Foundation of China (52222201, 52172032), Young Elite Scientist Sponsorship Program by CAST (YESS20200222), Hunan Natural Science Foundation (2022JJ20044), Shenzhen Science and Technology Innovation Commission Project (KQTD20221101115627004), and National University of Defense Technology (ZZCX-ZZGC-01-07). K.X. acknowledges support from the National Science and Technology Major Project (2023ZD0121101). Z.L. acknowledges support from the State Administration of Science, Technology, and Industry for National Defense (WDZC20245250509).

## Author Contribution

S.W., and K.X. initiated the project and generated the experimental protocols. Z.L. was responsible for data analysis and manuscript writing. M.F. and Z.G. wrote the code. S.W. prepared the samples and captured the experimental ADF-STEM and HR-TEM images. All authors contributed to the data analysis, manuscript writing, and manuscript revision.

## Conflict of interest

The authors declare no competing interests.

# Supplementary information

# Exploring structure diversity in atomic resolution microscopy with graph neural networks

*Zheng Luo[1,8], Ming Feng[2,3,8], Zijian Gao[2,8], Jinyang Yu[2], Liang Hu[3], Tao Wang[4], Shenao Xue[1,5], Shen Zhou[6], Fangping Ouyang[5], Dawei Feng[2], Kele Xu[2]*& Shanshan Wang[1,7]**

[1]Department of Materials, Hunan Key Laboratory of Mechanism and Technology of Quantum Information, College of Aerospace Science and Engineering, National University of Defense Technology, Changsha 410000, China

[2]School of Computer, State Key Laboratory of Complex & Critical Software Environment, National University of Defense Technology, Changsha 410000, China

[3]College of Electronic and Information Engineering, Tongji University, Shanghai 201804, China

[4]School of Material Science and Engineering, Peking University, Beijing 100871, China

[5]School of Physics, Central South University, Changsha 410083, China

[6]College of Science, National University of Defense Technology, Changsha 410000, China

[7]School of Advanced Materials, Peking University, Shenzhen Graduate School, Shenzhen, Guangdong 518055, China

[8]These authors contributed equally.

*Corresponding authors: K.X.(xukelele@nudt.edu.cn); S.W.(wangshanshan08@nudt.edu.cn)

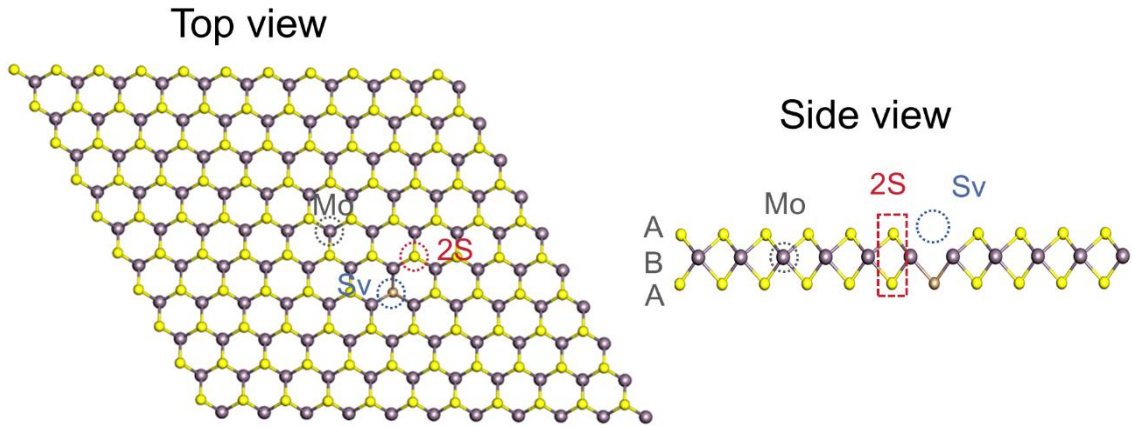

**Fig. S1.** Atomic model of monolayer MoS$_2$ containing a S vacancy. For monolayer 1H-MoS$_2$, the atoms in the S-Mo-S layers are stacked in an ABA sequence, where the S atoms in both the top and bottom layers occupy the same A sites and align along the vertical direction (square in red dashed line), while the Mo atoms (circle in black dashed line) occupy the B site. Therefore, we denote the atomic columns containing two aligned S atoms in the projection plane of pristine MoS$_2$ crystal as 2S, and the S vacancies with one S atom missing are denoted as S$_v$, which shows an intensity decrease compared to those 2S sites in the ADF-STEM image.

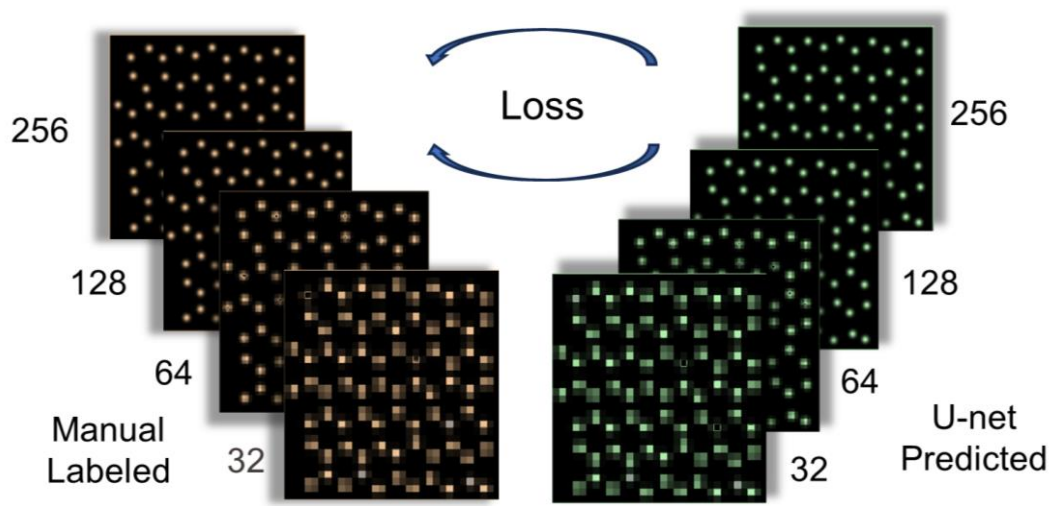

**Fig. S2.** Schematic showing the feature learning process of multi-scale U-Net through approximating the manual labels to the U-Net predictions via a loss function at different resolutions.

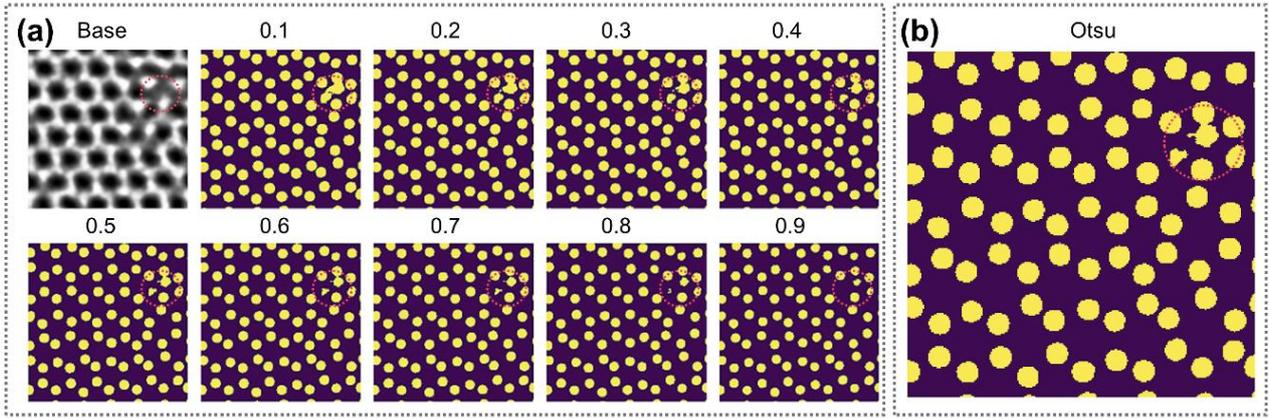

**Fig. S3.** Different threshold setting methods to divide the foreground and background in the generated probability maps of multi-scale U-Net. (a) Visualization of the probability map segmented via a manual threshold setting ranging from 0.1 to 0.9. (b) Probability map segmented with the self-adaptive Otsu algorithm. The test data employed here is an HR-TEM image of $MoS_2$ in a patch size of 256×256.

Generally, manually setting a fixed threshold is not always feasible as the segmentation threshold will be varied with the probability maps. As present in the red circle marks in Fig. S3a, if the preset threshold is too small, local details may be over-amplified while a large preset threshold may also lead to partial information loss, both of which would deteriorate the performance for accurately pinpointing the atoms. Otsu algorithm [1] can adaptively eliminate this bias as it enables to directly find the maximum between-class variance in each probability map to minimize the misclassification.

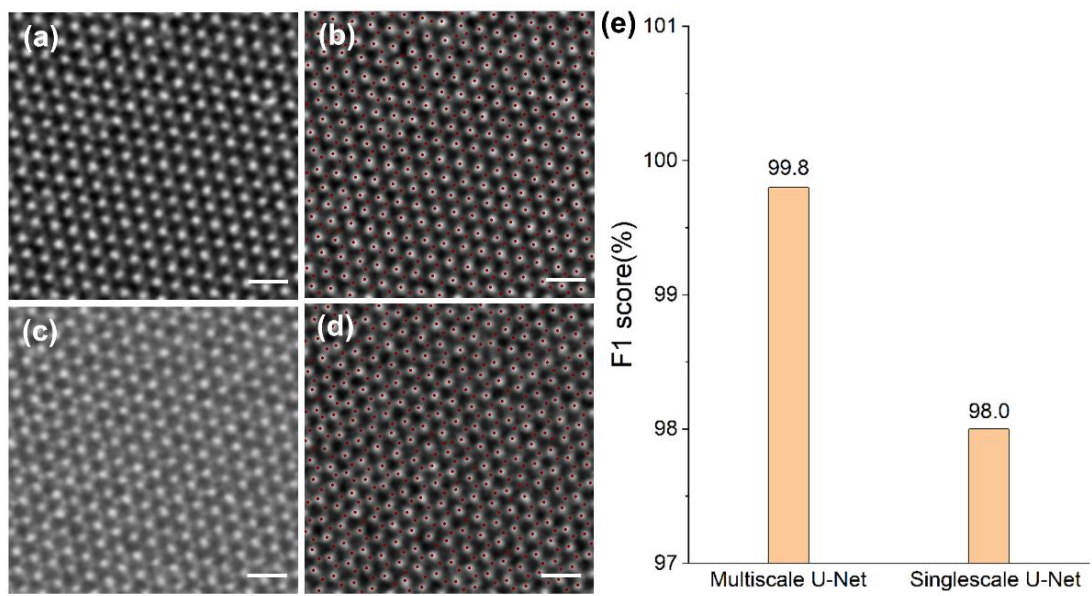

**Fig. S4.** Atomic localization results using our atom pinpointing frameworks. (a, b) ADF-STEM image of monolayer $MoS_2$ with astigmatism and its corresponding visualization result after pinpointing the atomic columns. (c, d) ADF-STEM image of monolayer $MoS_2$ with contamination and its corresponding visualization result after pinpointing the atomic columns. The red dots in (b) and (d) refer to the detected centroid of atomic columns. Scale bars: 0.5 nm. (e) F1 scores generated by multi-scale and vanilla U-Net for atom pinpointing. F1 scores are obtained by testing different models on 10 pieces of 512×512 ADF-STEM images involving ~5000 atoms.

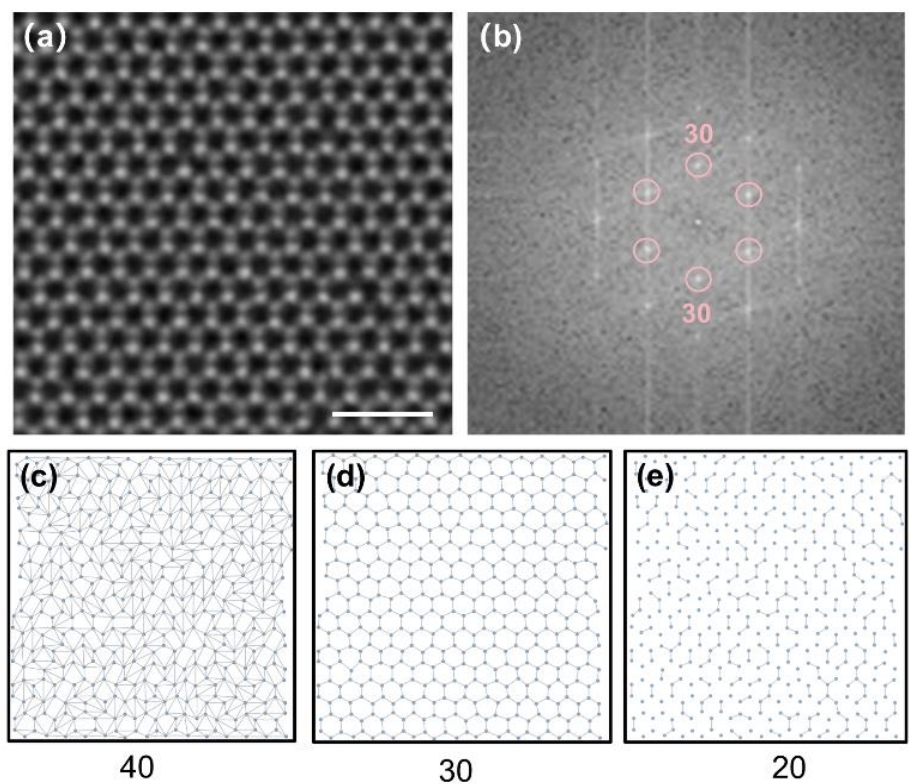

**Fig. S5.** Workflow to estimate the length threshold for edge connection by Delaunay triangulation. (a) ADF-STEM image of monolayer $MoS_2$. Scale bar: 1 nm. (b) FFT of the input image in (a). By searching the nearest spots (corresponding to the (001) plane of $MoS_2$, marked by pink circles) in the generated FFT, the pixel distance corresponding to the interplanar spacing in real space can be directly extracted from Image J software (the pixel distance is 30 in real space). The edge length that exceeds the threshold will be removed during the Delaunay triangulation process. (c-e) The connection results in a length threshold of 40, 30, and 20, respectively. The overcounting or disconnection issues are obvious if the threshold is immoderate.

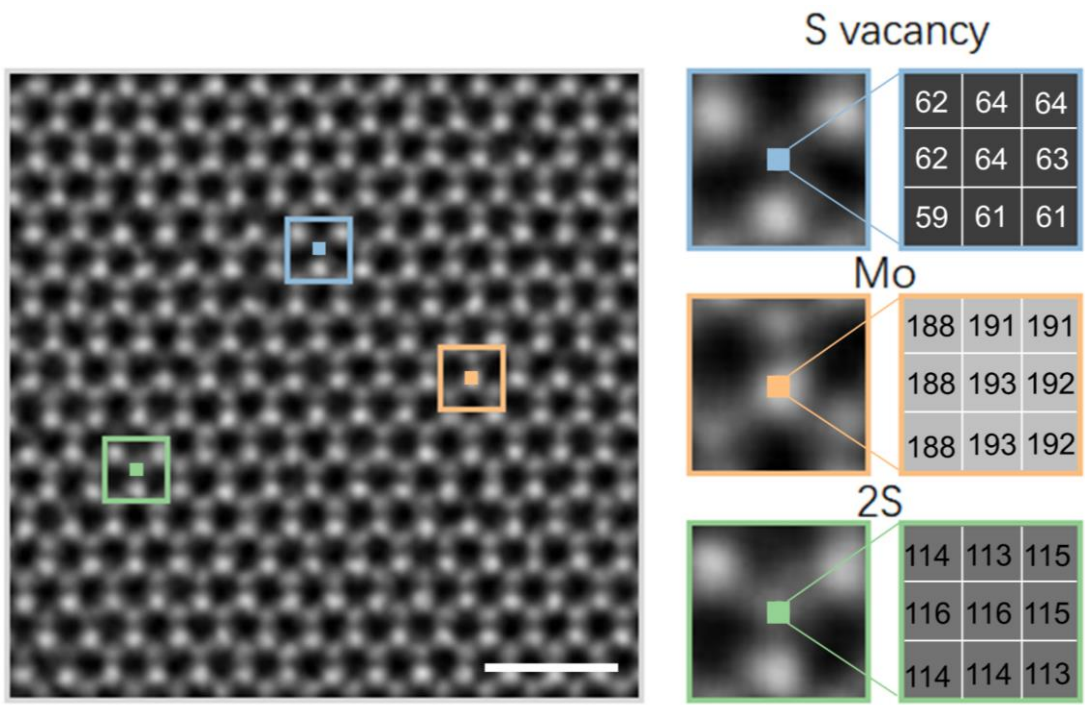

**Fig. S6.** ADF-STEM image of monolayer $MoS_2$ with each node embedded in a 3×3 gray value matrix by taking nodes corresponding to S vacancy (blue dot), Mo (orange dot), and 2S (green dot) as examples. Scale bar: 1 nm.

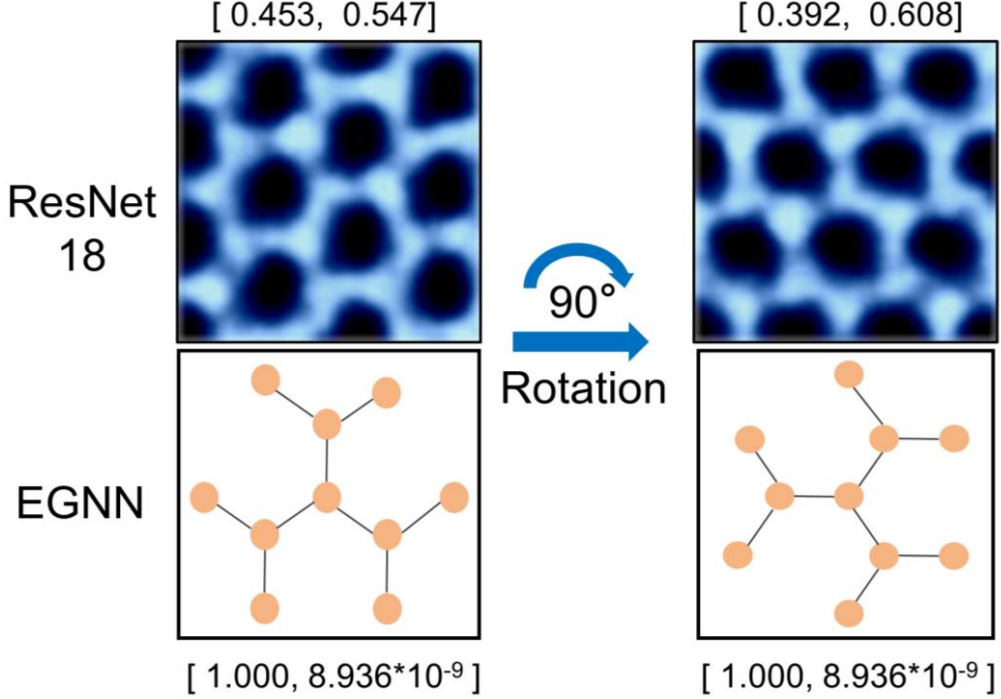

**Fig. S7.** Output probability for the classification of the center atomic column using ResNet 18 and EGNN models after the input image patch and graph sample experiencing 90° rotation. It is obvious that the output probability generated by ResNet 18 (the values in the brackets, where the left one corresponds to the probability of normal atomic column and the right one corresponds to the probability of vacancy) simultaneously changes with the rotation of the input image. On the contrary, the output probability from EGNN remains unchanged, demonstrating the robustness of the EGNN model against rotation variation.

In this section, we need to prove that the EGNN is rotation-equivariant on $x$ (spatial coordinates) for any orthogonal matrix $Q \in \mathbb{R}^{n \times n}$. We need to prove that the model satisfies:

$$Qx^{l+1}, h^{l+1} = EGCL(Qx^l, h^l) \tag{S1}$$

Assuming $h^0$ is invariant to $x$'s rotation transformation, then the output $m_{ij}$ from Equation 3 (see details in **Methods**) will also remain invariant to the rotation $\| Qx_i^l - Qx_j^l \|^2 = (x_i^l - x_j^l)^\top Q^\top Q (x_i^l - x_j^l) = (x_i^l - x_j^l)^\top I (x_i^l - x_j^l) = \| x_i^l - x_j^l \|^2$, making the edge operation invariant:

$$m_{i,j} = \phi_e \left( h_i^l, h_j^l, \| Qx_i^l - Qx_j^l \|^2 \right) = \phi_e \left( h_i^l, h_j^l, \| x_i^l - x_j^l \|^2 \right) \tag{S2}$$

The equation for updating the coordinates $x$ in the model is rotation-equivariant. Its equivariance can be proved if the rotation transformation of the input leads to the same transformation in the output. Note that, as shown in the proof above, $m_{ij}$ is already invariant.

$$Qx_i^{l+1} = Qx_i^l + \frac{1}{M-1}\sum_{j \neq i}(Qx_i^l - Qx_j^l)\phi_x(m_{i,j}) \tag{S3}$$

Differentiating this equation, we get:

$$\begin{aligned}Qx_i^l + \frac{1}{M-1}\sum_{j \neq i}(Qx_i^l - Qx_j^l)\phi_x(m_{i,j}) &= Qx_i^l + Q\frac{1}{M-1}\sum_{j \neq i}(x_i^l - x_j^l)\phi_x(m_{i,j}) \\ &= Q\left(x_i^l + \frac{1}{M-1}\sum_{j \neq i}(x_i^l - x_j^l)\phi_x(m_{i,j})\right) \\ &= Qx_i^{l+1}\end{aligned} \tag{S4}$$

Therefore, rotation on $x^l$ will result in the same rotation on $x^{l+1}$ from the Equation 4's output (see details in **Methods**). Furthermore, Equations 5 and 6 (see details in **Methods**) only depend on $m_{ij}$ and $h^l$, which are rotation-invariant shown in the beginning of the proof. Thus, the output of Equation 3 (see details in **Methods**), $h^{l+1}$, will also be invariant. Consequently, we conclude that the transformation $Qx^{l+1}, h^{l+1} = \text{EGCL}(Qx^l, h^l)$ is satisfied.

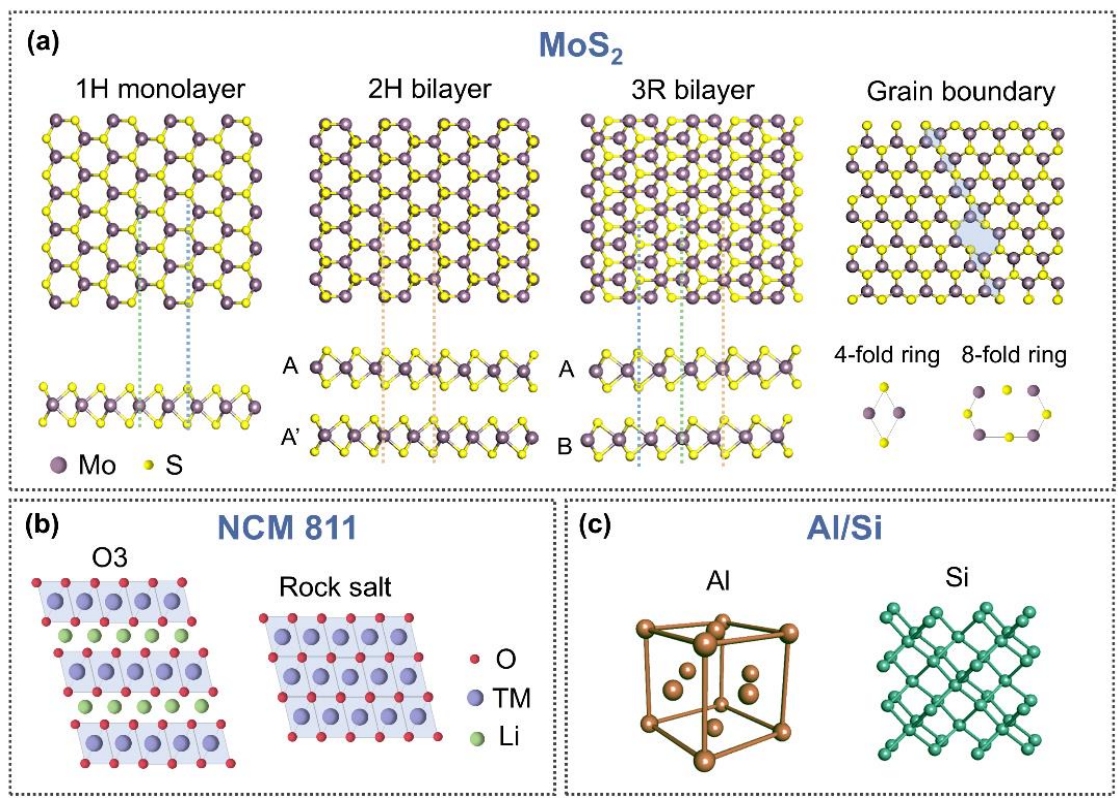

**Fig. S8.** (a) Atomic models showing the stacking configurations of 1H (monolayer), 2H (bilayer) and 3R (bilayer), aas well as the grain boundaries in monolayer MoS₂. The orange, blue, and green dashed lines align the Mo+2S, 2S, and Mo in the sectional views with their corresponding atomic columns in the projection plane. The blue area shows the grain boundary composed of 4- and 8-fold rings. (b) Atomic models of O3 and rock salt phase in a NCM 811 cathode material of lithium-ion battery. (c) Atomic models of aluminum and silicon.

$MoS_2$ is a typical two-dimensional transition metal disulfide, which generally presents a thermodynamically stable hexagonal structure [2]. *1H* monolayer $MoS_2$ has Mo atoms bonded to S atoms in a trigonal prismatic coordination by strong covalent bonds (the first panel in Fig. S8a), while the interlayer is bonded with a weak van der Waals (vdW) force, which gives rise to polytypism in bilayer $MoS_2$ with different stacking structures. Typically, the *2H* and *3R* polytypes of $MoS_2$ bilayers in AA′ and AB stacking are the most energetically stable and commonly observed stacking configurations in synthetic materials. For *2H* stacking (the second panel in Fig. S8a), the Mo (S) atoms in one layer overlap with the S (Mo) atoms in the other layer (orange dashed line), maintaining a hexagonal structure. 3R stacking has double stacked S atoms in one layer that are located above (below) the centers of the hexagonal rings in the other layer (the third panel in Fig. S8a), exhibiting a rhombohedral structure. In this configuration, three different atomic columns are observed in the top-down projection, a single Mo atom, two S atoms plus one

Mo atom, and two S atoms, respectively. Stacking is one of the key attributes of layered MoS$_2$, where the number of layers and their stacking arrangement largely modify its electronic properties. For example, the *1H*-MoS$_2$ exhibits a direct band gap in contrast to the indirect band gaps observed in the *2H* and *3R* stacking configurations [3]. Moreover, highly localized metallic states have been demonstrated at the atomically sharp *2H-3R* stacking boundaries [4]. Therefore, the rapid identification of each stacking configuration and the stacking boundaries is crucial for understanding the mechanism behind stacking formation and designing advanced nanodevices.

Grain boundary is a typical line defect in MoS$_2$, which is composed of different dislocation cores (like the 4- and 8-fold rings present in the fourth panel in Fig. S8a), enabling a flexible property modulation (like mechanical and electrical properties) for pristine MoS$_2$ [5]. However, due to the nanoscale width of grain boundaries, directly locating them on monolayer MoS$_2$ at the microscopic level using ADF-STEM is time-consuming and challenging. Our framework enables rapid detection of grain boundaries and detailed investigation into their atomic configurations (like the composition of 4- and 8-fold rings), which helps reveal the underlying structure-property correlation in polycrystal MoS$_2$.

The O3 phase in NCM 811 cathode material exhibits a rhombohedral structure (left panel in Fig. S8b), where the transition metal atoms (Ni, Co, Mn, abbreviated as TM) and the oxygen atoms jointly constitute a stable TMO$_2$ layer to create interlayer transportation channels for reversible Li$^+$ intercalation/deintercalation. However, severe cation mixing (between Li and Ni) will be generated in the NCM 811 cathode material during battery cycling, which ultimately forms an inactive rock salt phase (cubic structure shown in the right panel in Fig. S8b)), leading to the rapid capacity decay [6]. With the assistance of our framework, the region from which the phase transition originates can be automatically identified, thereby providing a robust platform for comprehending the underlying mechanism governing battery capacity degradation at an atomic level.

The crystal structure of aluminum belongs to a typical face-centered cubic lattice (left panel in Fig. S8c). Silicon has a crystal structure similar to diamond, where each silicon atom is connected with the other four atoms through covalent bonds to form a tetrahedral network. (right panel in Fig. S8c). The aluminum plating on silicon wafers is widely used to create metal wires or contact electrodes in electronic devices or integrated circuits, which should form close contact with the silicon substrate to achieve stable current transmission [7]. Our framework provides convenience to rapidly locate the hetero-interfaces for contact quality evaluation in semiconductor manufacturing.

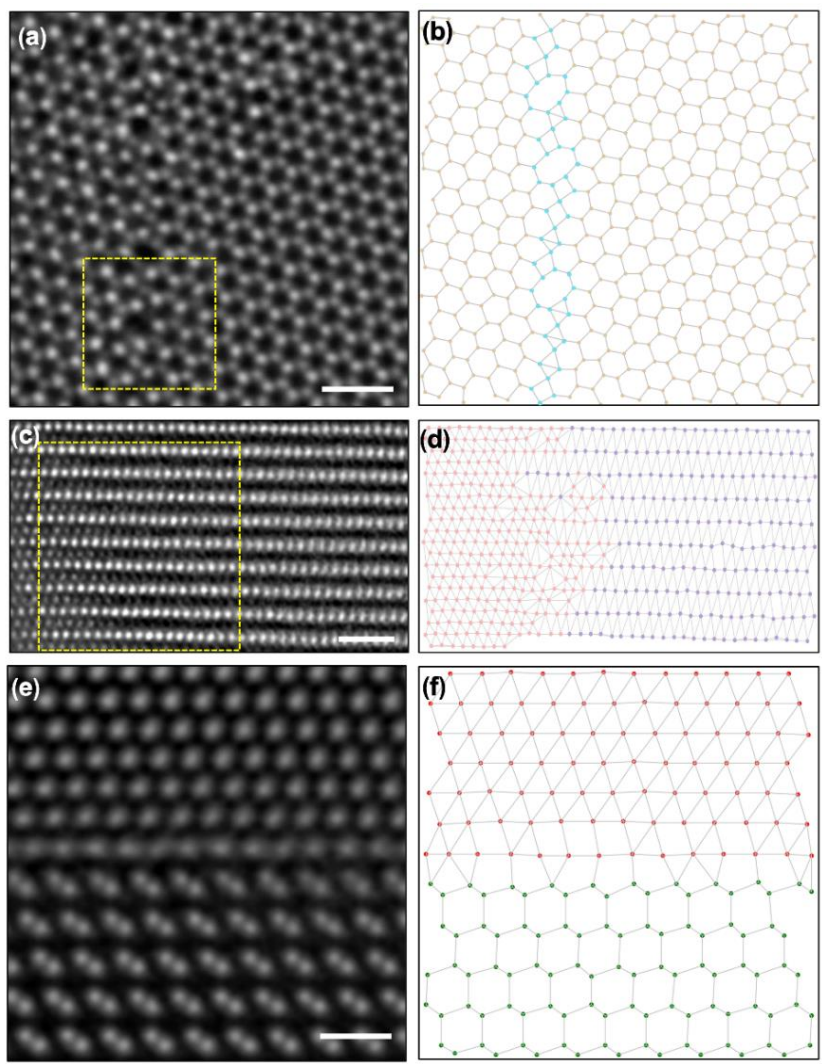

**Fig. S9.** Identification of boundaries and interfaces among different material systems. (a) ADF-STEM image of monolayer $MoS_2$ (512×512) containing a grain boundary composed of different dislocation cores. Scale bar: 1 nm. (b) Identification result of grain boundary in the form of a planar graph representation. The polygonal rings composed of cyan dots show the identified dislocation cores in (b). (c) ADF-STEM image of layered cathode material containing rock salt and O3 phase (1024×512). Scale bar: 0.5 nm. (d) Identification result of phase boundary in the form of a planar graph representation. The pink and purple dots in (d) correspond to the identified rock salt and O3 phases. (e) HR-TEM image of a hetero interface between Al and Si (512×512). Scale bar: 0.5 nm. (f) Identification result of the hetero interface in the form of a planar graph representation. The red and green dots in (f) correspond to the identified Al and Si phases. The squares in the yellow dashed line in (a) and (c) show the corresponding cropped image patches for Fig. 3a (ii) and (iii), respectively. Note that, the identification of grain boundary was achieved by incorporating the graph-level feature via searching the minimum closed loops from the as-generated planar graphs. The other two tasks were achieved by incorporating the node-level feature. All these models were trained by only one piece of different structure corresponded image (512×512).

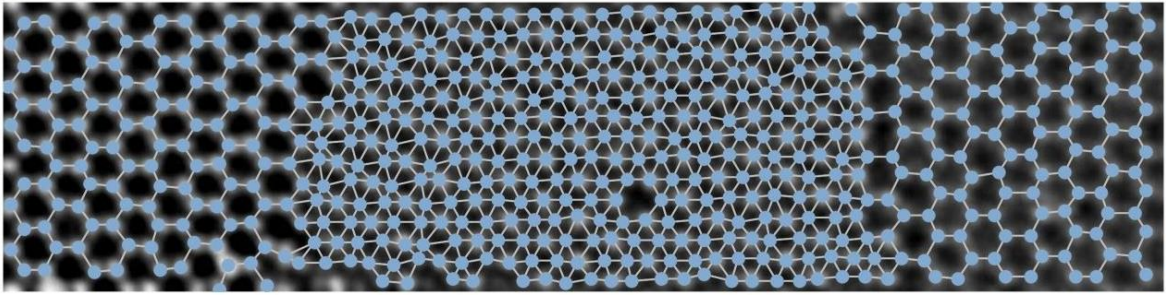

**Fig. S10.** The as-generated planar graph using Delaunay triangulation.

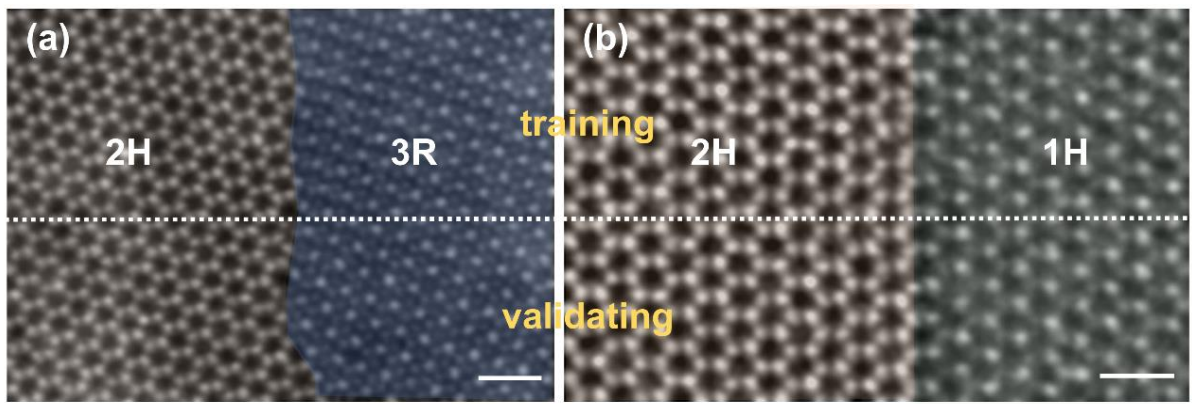

**Fig. S11.** Datasets for EGNN model training. We chose two ADF-STEM images containing different stacking configurations of *2H-3R* (a) and *2H-1H* (b) to learn their graph feature, where the upper half of these two images were employed for model training and the lower parts were employed for validating. Scale bars: 0.5 nm.

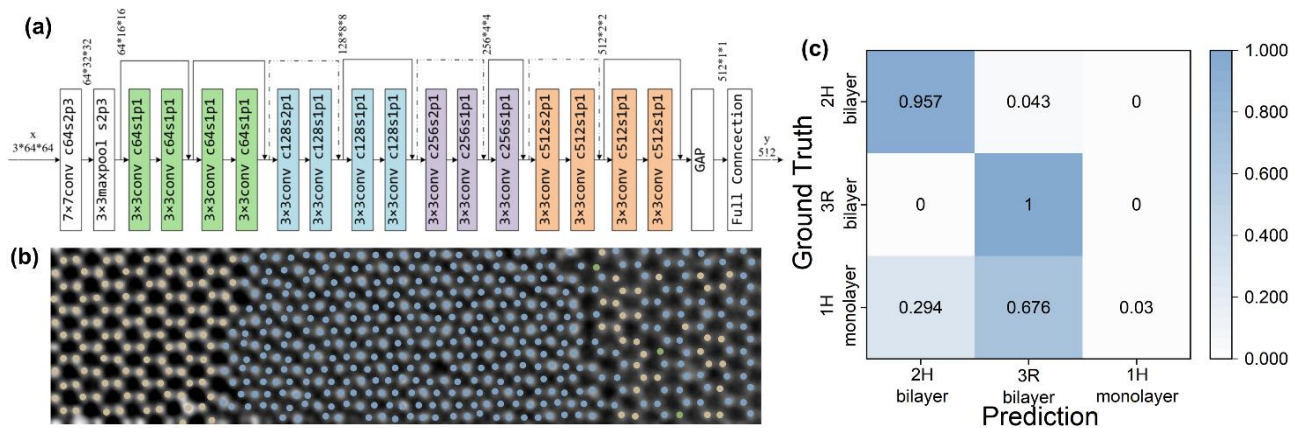

**Fig. S12.** The segmentation result of different stacking configurations from Fig. 3b using a typical CNN model (ResNet 18). (a) Frameworks of ResNet 18. (b) End-to-end visualization result obtained by ResNet 18. (c) Confusion matrix generated by EGNN model in the segmentation of different stacking configurations from Fig. 3b.

ResNet 18 is a state-of-the-art deep convolutional neural network [8] that incorporates 18 layers of residual blocks to facilitate the training of deeper architectures (Fig. S12a). In this stacking configuration segmentation task, patches were extracted in size of 64×64 with any identified atom as the center. From the confusion matrix (Fig. S12c), the model trained by ResNet 18 exhibits extremely poor performance in the classification of 1H monolayer and 3R bilayer (Fig. S12b) compared to that using EGNN mode, showcasing the over-fitting phenomena of image-based learning framework in a few-shot training dataset.

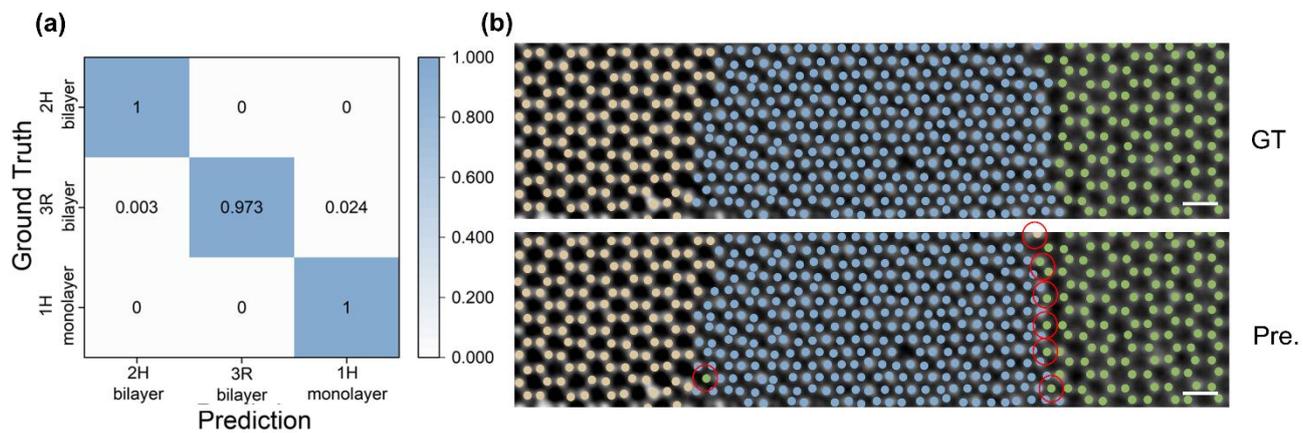

**Fig. S13.** (a) Confusion matrix generated by our framework from Fig. 3b. (b) Detailed investigation of the misidentified points in the segmentation of different stacking configurations predicted by our framework (marked by red circles). Scale bars: 0.5 nm. The abbreviations of GT and Pre. refer to the ground truth and the prediction.

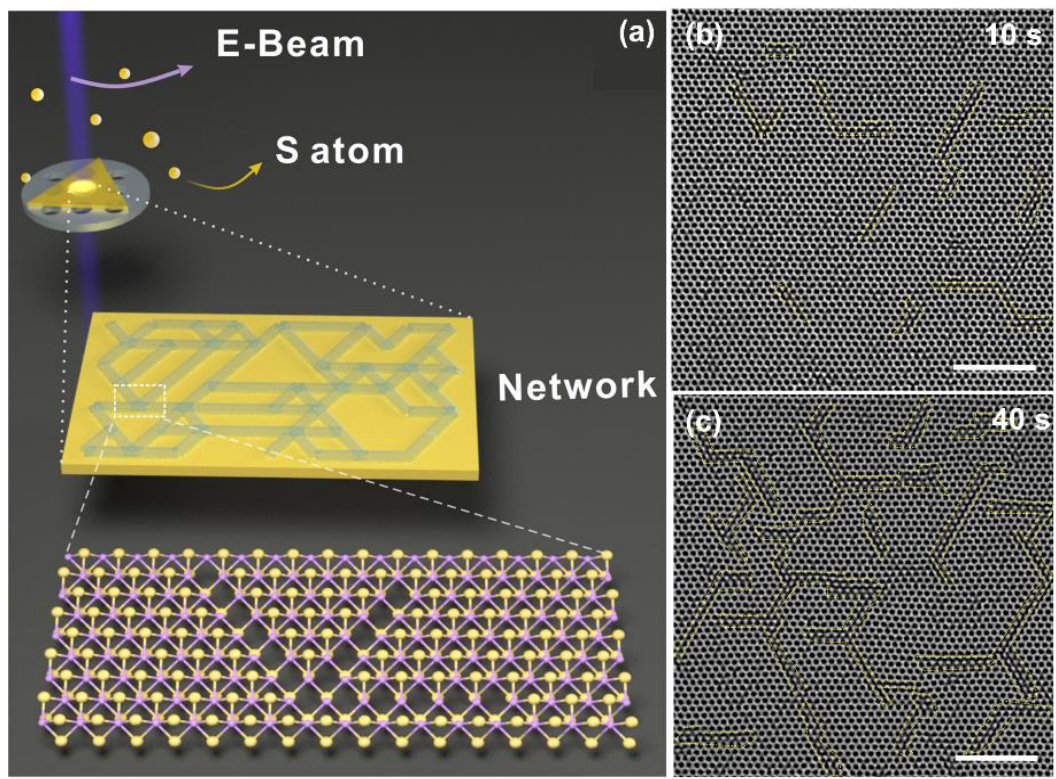

**Fig. S14.** Evolution of the aggregated vacancy lines under continuous electron beam irradiation in monolayer MoS$_2$. (a) Schematic showing the generation process of vacancy lines. (b) HR-TEM image of monolayer MoS$_2$ at the initial state. (c) HR-TEM image of monolayer MoS$_2$ after 40 s electron beam irradiation. Scale bars: 4 nm.

Beyond imaging, electron beam irradiation with either a "knock-on" effect, ionization, or beam-induced chemical etching effects, has been demonstrated to induce structural disorder at the atomic level, facilitating the atomic sculpturing of 2D membranes with high spatial accuracy and flexibility. In this work, a series of in-situ HR-TEM images were captured from a fast frame rate camera with an accelerating voltage of 80 kV, where only S rather than Mo atoms can be ejected under this irradiation condition, manifesting intensity decrease and blurring compared to 2S site in the HR-TEM images [9].

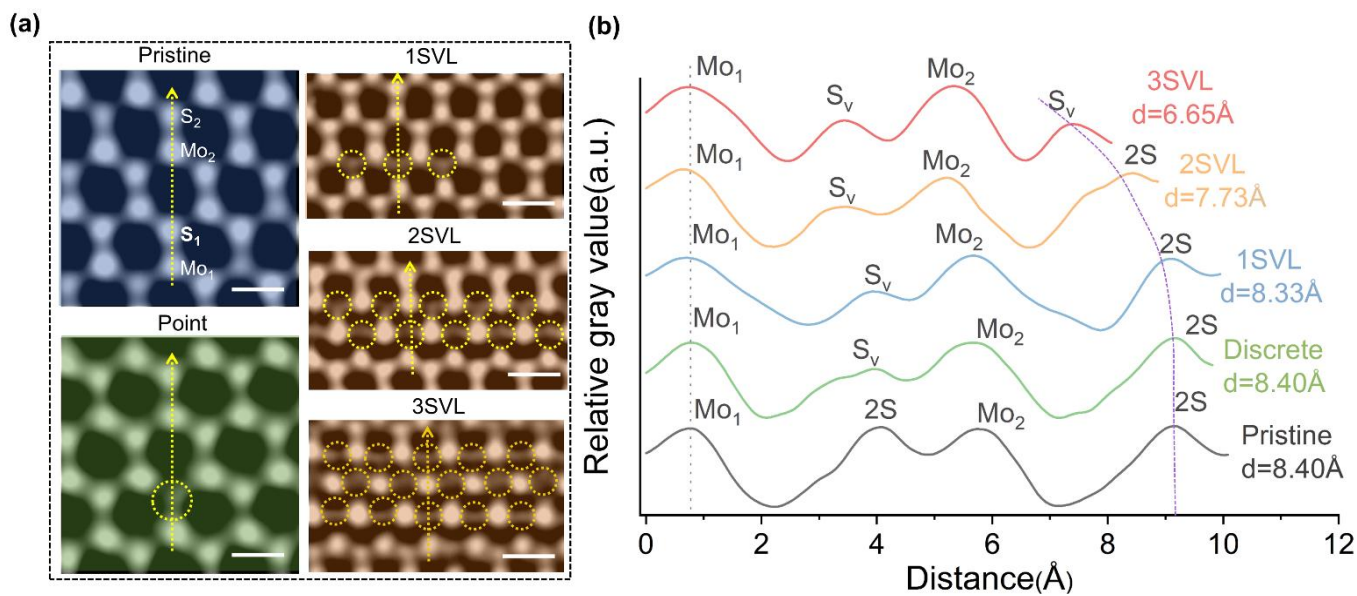

**Fig. S15.** (a) HR-TEM image patches of pristine structure, discrete vacancy, and different morphology of vacancy lines (1SVL, 2SVL, 3SVL). The arrows in each image patch indicate the armchair direction from $Mo_1$ to $S_2$ and the dashed yellow circles refer to the S vacancies. Scale bars: 0.25 nm. (b) Intensity line profiles show the distance variation along the armchair direction. The interatomic distance of $Mo_1$-$S_2$ along the armchair direction continuously decreases as the line defect broadens (the purple dashed line in Fig. S15b shows the variation of $S_2$ sites), and finally decreases from 8.40 Å to 6.65 Å with the formation of 3SVL (nSVL is denoted as the number of adjoining parallel S vacancy lines), yielding a shrinkage of 20.8%. On the contrary, the interatomic distance of $Mo_1$-$S_2$ along the armchair direction remains almost unchanged with the formation of discrete vacancies.

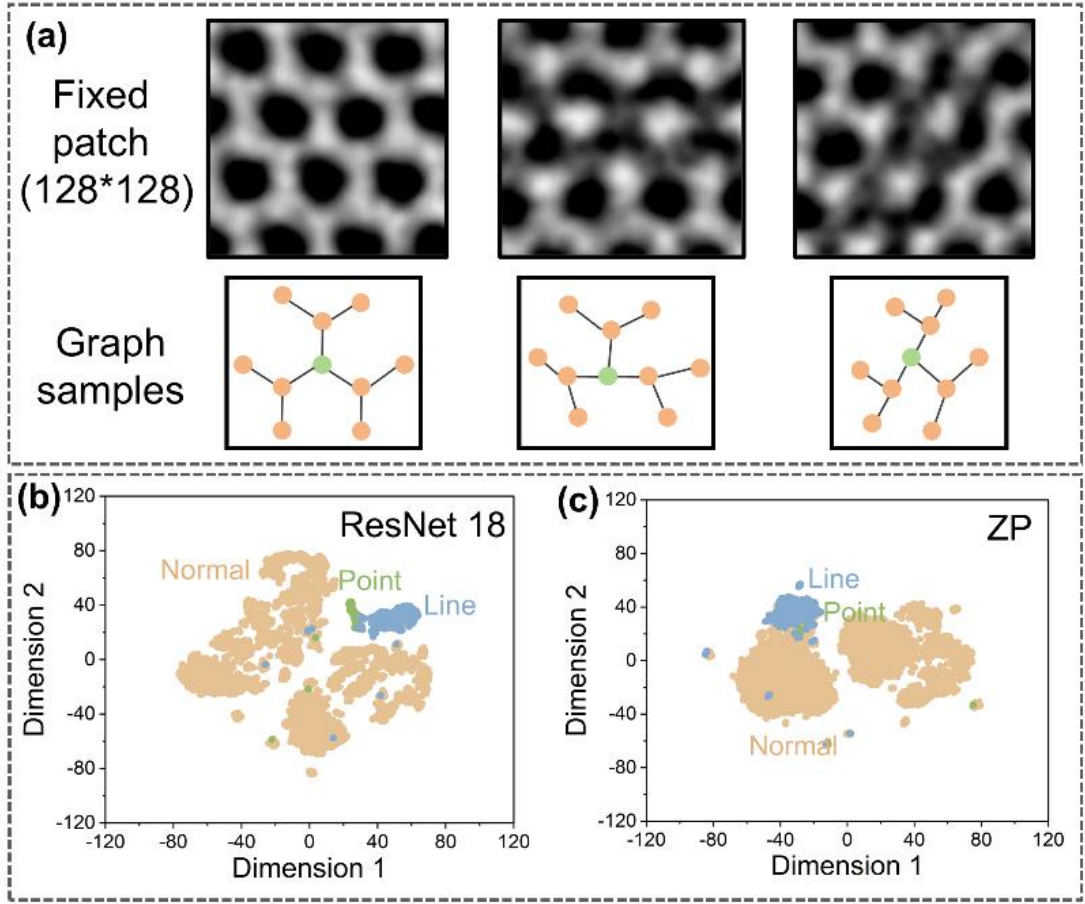

**Fig. S16.** (a) Feature extraction by CNN, Zernike polynomials (ZP) from fixed patches, as well as by EGNN from graph representation. (b, c) T-SNE outputs in feature space generated by ResNet 18 and ZP.

The ZP is a complete set of orthogonal basis functions denoted by the double indexing sheme $Z_q^p(\rho, \theta)$ which can decompose image patch to a linear combination of ZP. The coefficients for each polynomial can be grouped to form a compact representation of the original patch, which has been employed by Dan et al. [10] to extract motif features of microscopic structures, showing impressive performance in clustering several ordered structures like doping and vacancies. However, ZP extracts features in a fixed image patch. When the patch size is inappropriate, the local environment information will either be lost or share high similarity with the others, which has been demonstrated to deteriorate the feature extraction ability of ZP in Dan's work. Therefore, when processing vacancy lines with flexible levels of lattice distortion, the local structure variation makes it difficult to effectively extract features of the aggregated line and point vacancies in a fixed image patch size, resulting in the fuzzy boundary of different clusters in feature space (Fig. S16c).

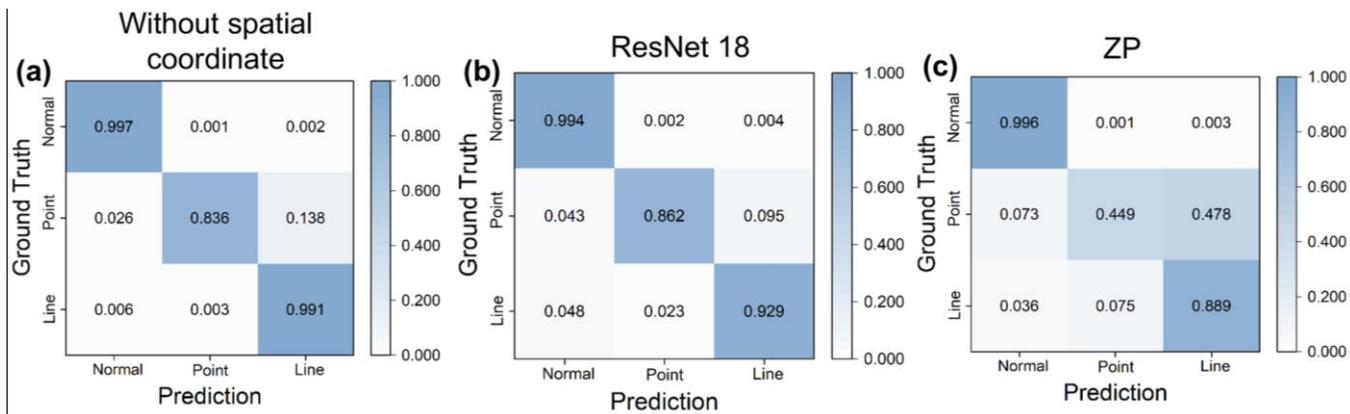

**Fig. S17.** (a) Confusion matrix generated by the EGNN model without embedding spatial coordinate information in each node. (b) Confusion matrix generated by ResNet 18 model. (c) Confusion matrix generated by ZP. All these results are evaluated from Fig. 4a.

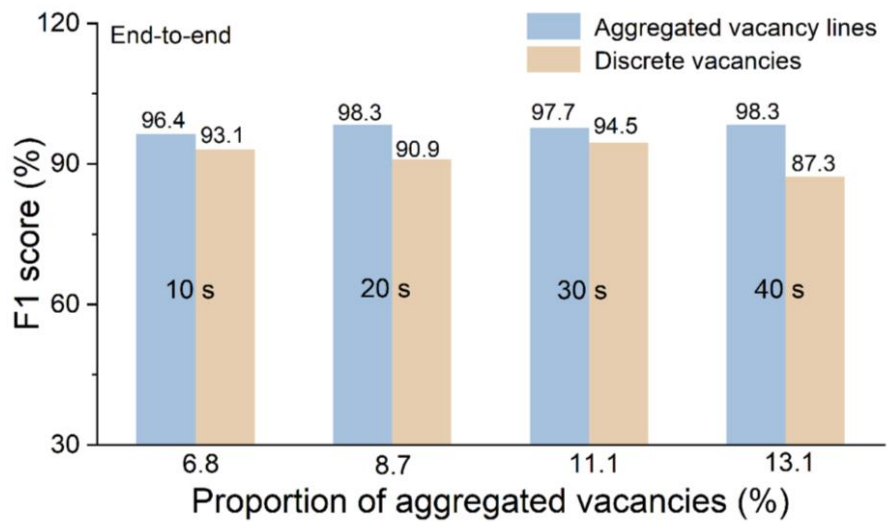

**Fig. S18.** F1 scores of the EGNN model test on the samples with different proportions of aggregated vacancies.

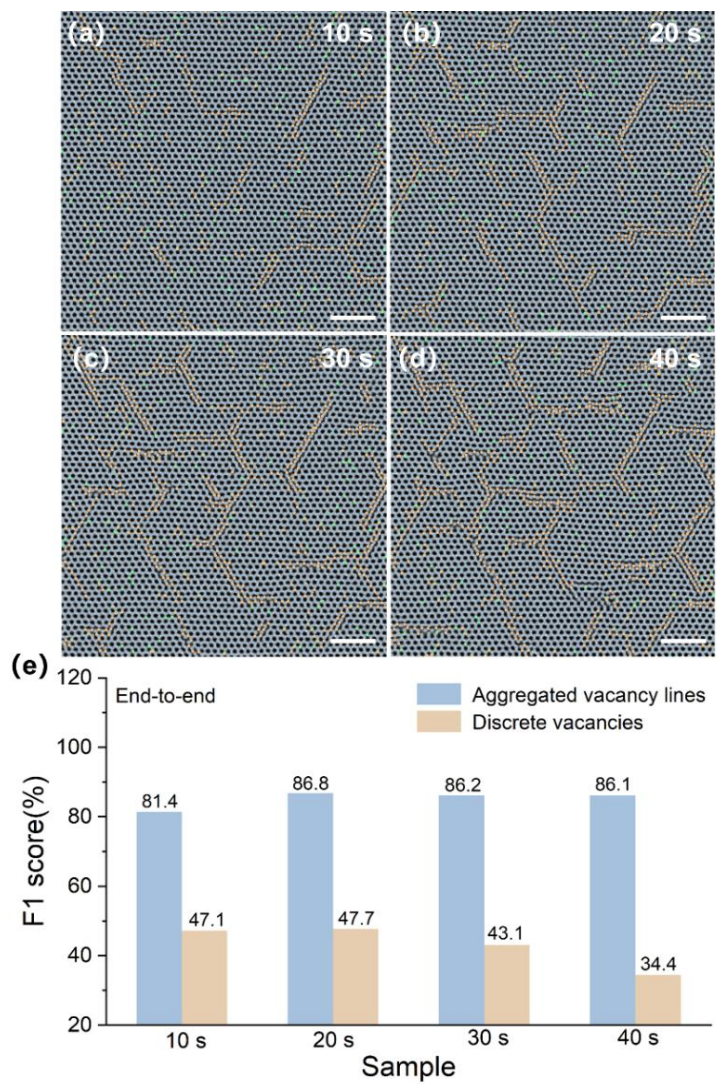

**Fig. S19** (a-d) End-to-end recognition results of the monolayer MoS$_2$ under different irradiation times using muti-scale U-Net framework alone. Scale bars: 2 nm. (e) F1 scores of multi-scale U-Net for the classification of aggregated and discrete vacancies.

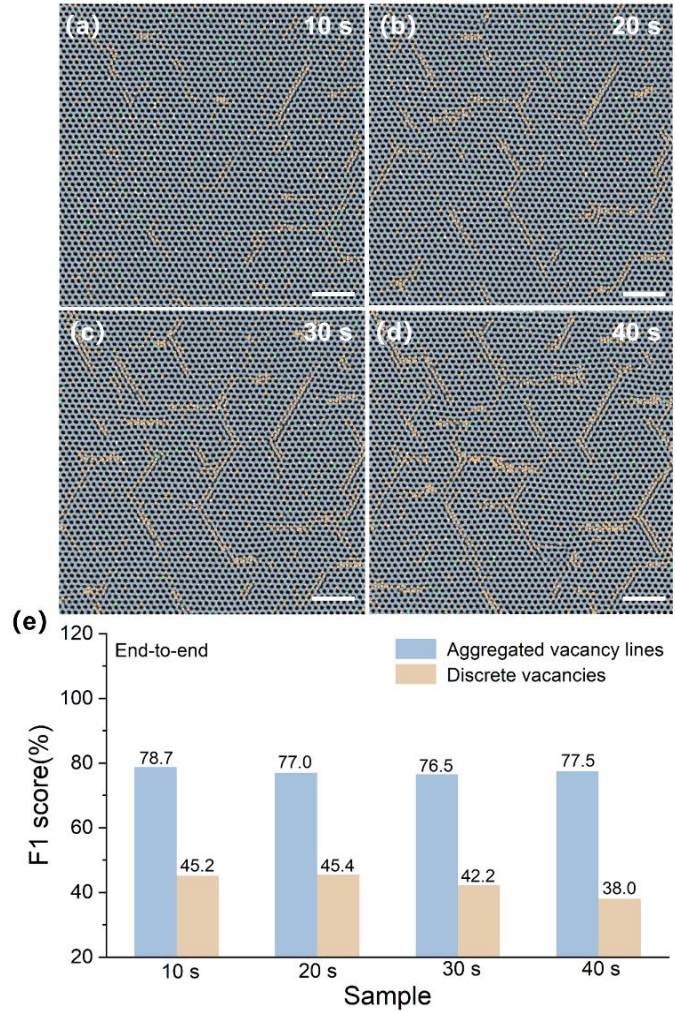

**Fig. S20.** (a-d) End-to-end recognition results of the monolayer MoS$_2$ under different irradiation times using the YOLO framework. Scale bars: 2 nm. (e) F1 scores of the YOLO model for the classification of aggregated and discrete vacancies.

The YOLO series is the most widely used object detection algorithm with fast speed and remarkable performance, where the bounding box and class of an object are predicted after a single forward process [11]. As a comparison of our work, we detect the vacancies based on the YOLOv5-m model. We introduced HR-TEM crops in size 256×256 with 50% overlaps and resized them to the regular size of 640×640. Each crop has several atoms with different classes as objects, with the bounding box size of 20×20 centered around the atom coordinate. We also employed basic data augmentation, such as random flipping, rotation, mosaic, and mix-up. From the identification result (Fig. S20e), the model trained by YOLOv5-m underperformed in the separation of aggregated line and discrete vacancies, further verifying the limitation of pixel-based DL framework for the identification of defects with a flexible range of lattice distortions in fixed image patches.

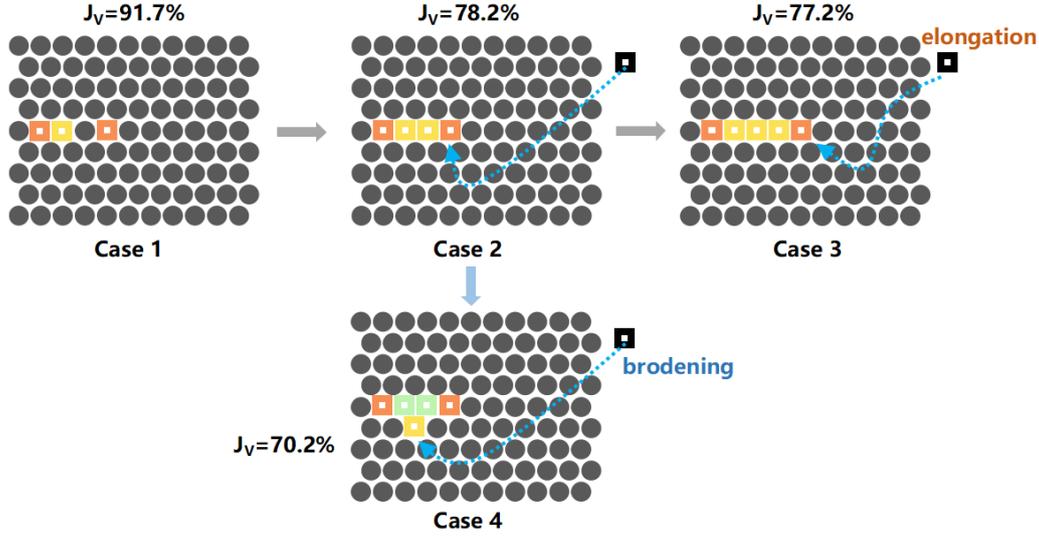

**Fig. S21.** Schematic model showing the relation between the alloying degree $J_v$ and the topology evolution of the vacancy aggregates.

The 'alloy degree' has been extensively utilized to characterize the atomic-scale mixing state of the components in a binary alloy system [12]. In this study case, we consider the vacancy as a special type of alloy element, and the sum of different coordinated vacancies gives the value of the alloying degree '$J_v$' as:

$$J_v = \frac{P_{observed\text{-}S}}{P_{random\text{-}S}} \tag{S5}$$

$$P_{observed\text{-}S} = \frac{\sum_{i=0}^{6}(i \times N_{v-iS})}{6 \times N_V} \tag{S6}$$

$$P_{random\text{-}S} = \frac{N_S}{N_S + N_v} \tag{S7}$$

where $P_{observed\text{-}S}$ is defined as the ratio of the averaged 2S coordination number around a vacancy to the total coordination number (six), and $P_{random\text{-}S}$ is the atomic ratio of 2S in the observed area, $N_v$ and $N_S$ refer to the total number of the $S_v$ and the 2S in the examined area, and $N_{v-iS}$ indicates the number of the $S_v$ with $i$ (0-6) coordinated 2S atomic columns around. Theoretically, if there is no preference for the arrangement of $S_v$ and 2S, $J_v$ should be 100%. If $J_v$<100%, both 2S and $S_v$ prefer to aggregate with themselves to form a homophilic configuration. If $J_v$>100%, S vacancies are prone to gather disulfide sites around and form a heterophilic configuration.

The growth anisotropy of vacancies results in different aggregation morphology, which can be mathematically reflected by evaluating the decrease rate of $J_v$ from a model containing 99 atoms with different vacancy distributions. As present in Fig. S21, the $J_v$ exhibits a rapid decrease from 91.7% to 78.2% (decreased by 13.5%) when a new vacancy is added to connect the originally discrete vacancies to a line (case 1 to case 2), which is well aligned with the first round of rapid $J_v$ reduction in Fig. 4h, indicating the morphology change of vacancies from isolated clusters to aggregated lines. Further prolonging the vacancy line by adding additional vacancy along the zigzag direction (case 2 to case 3) induces a minor $J_v$ decrease from 78.2% to 77.2% (decreased by 1.0%), while broadening the vacancy line along the armchair direction (case 2 to case 4) results in a moderate $J_v$ decrease from 78.2% to 70.2% (decreased by 8.0%). Therefore, the platform and the moderately declined slash from Fig. 4h in the later stage therefore can be attributed to the growth in length and width of the vacancy lines, respectively.

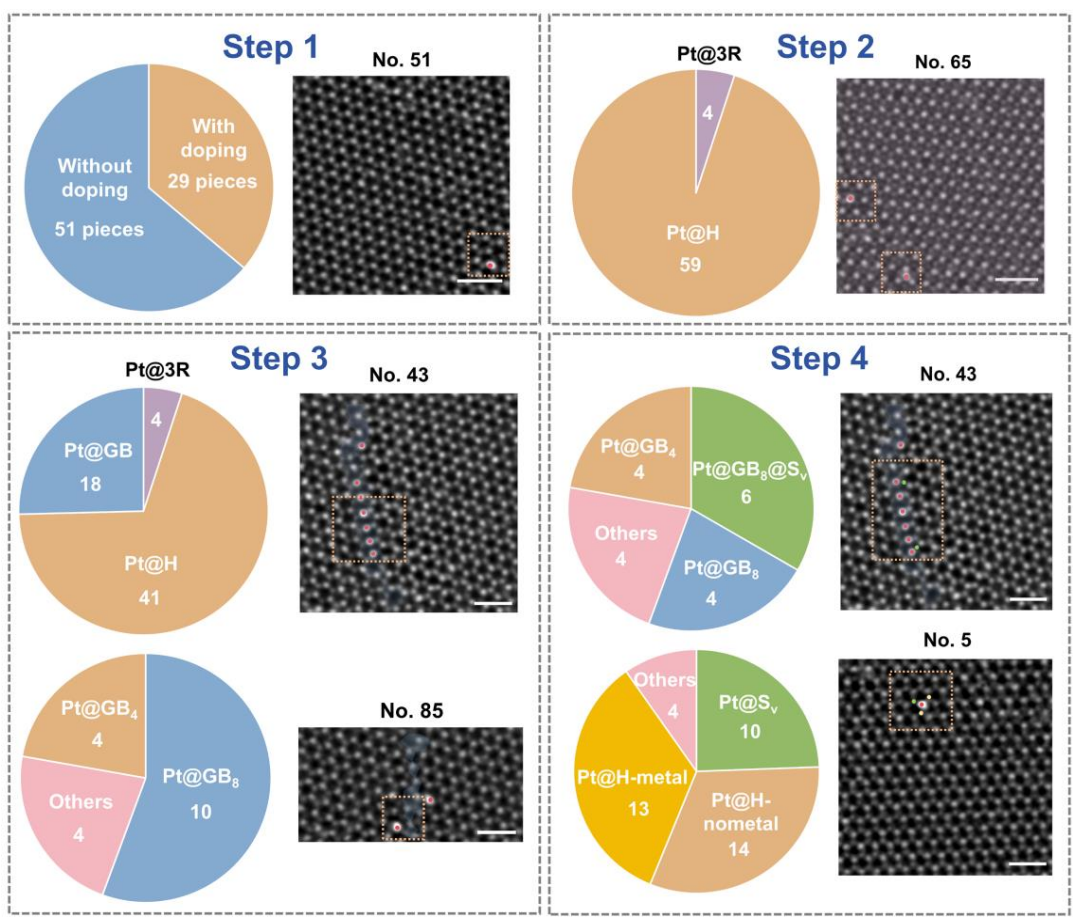

**Fig. S22.** Statistical results obtained by a step-by-step analysis of the outputs from different sub-models. Our task chain was employed to analyze a dataset containing 80 pieces of ADF-STEM images (512×512). Each image with a serial number from No.1 to No.80. The ADF-STEM images next to each pie chart are the typical identification generated at different steps. The orange square in the dashed line highlights the specific identification results in each step. Scale bars: 0.5 nm. In the first step, we employed the EGNN sub-model corresponding to doping detection to screen out the images containing Pt doping atoms (29 pieces of ADF-STEM images), and the total number of Pt atoms was calculated to be 63. In the second step, we employed the EGNN sub-model corresponding to the stacking configuration to analyze the Pt distribution in different stacking configurations, where four Pt atoms were found to be situated in the 3R stacking by overlaying the recognition results of the 3R stacking or 1H monolayer with that of the doping. In the third step, the images containing Pt situated in the 1H monolayer were analyzed by the EGNN sub-model corresponding to the grain boundary detection. After overlaying the outputs with that of the doping, 18 Pt atoms were found to be situated in the grain boundaries. Furthermore, by counting the number of Pt atoms on different minimum closed loops, it can be found that the Pt atoms are primarily situated in the 8-fold rings of grain boundaries. In the final step, we employed the sub-model for the identification of vacancies among the images containing Pt atoms. By searching the nearest neighbors of Pt atoms that contain vacancies and overlaying these results with that of the stacking configurations and grain boundaries, 10 Pt@$S_v$ and 6 Pt@$GB_8$@$S_v$ configurations were screened out, which were subsequently employed for DFT calculations. In addition, this process can be further extended to investigate the detailed coordinate environment of each Pt atom in the Pt@1H configurations, by searching the number of the nearest identified Mo and 2S atomic columns centered around Pt atoms, where 14 and 13 Pt are found to be situated in the nonmetal and metal site of the Pt@1H configurations. Note that, several misidentifications like the Pt situated in 5-fold rings (others) of grain boundaries were detected when tracing back to their raw images. However, these misidentifications would not significantly affect the discovery of new doping configurations.

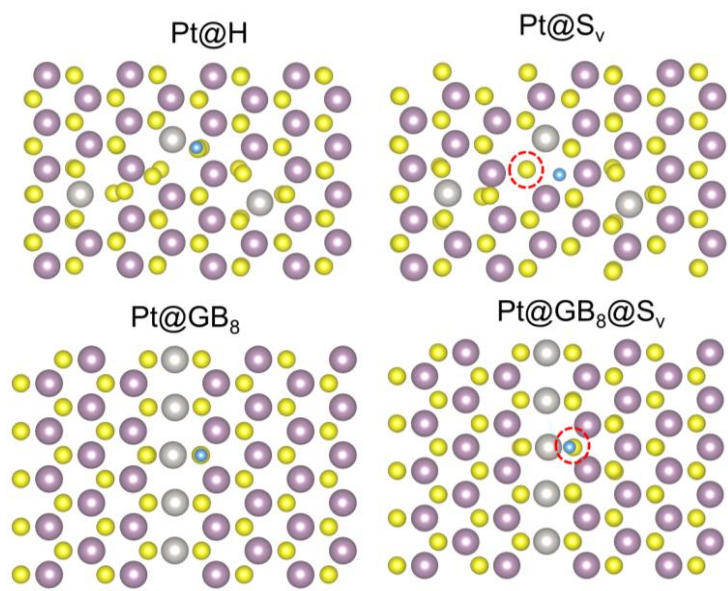

**Fig. S23.** H* adsorption sites on different Pt doping configurations after structural optimization. The purple, yellow, gray, and blue balls represent the Mo, S, Pt, and H* atoms, respectively. The red circles in the dashed line show the S vacancy sites.

**Table S1** Graphics memory cost and computing parameter of different models in the phase segmentation task (For ResNet 18 model training, the patch size is 64×64, the batch size is 32).

| Model | Time cost (s) | Graphics memory cost (MB) | Computing Parameter (M) | FLOPs (M) |
|---|---|---|---|---|
| ResNet 18 | $2.7×10^{-2}$ | $6.7×10^{2}$ | 11.6 | $1.5×10^{5}$ |
| EGNN | $6.0×10^{-3}$ | $3.4×10^{2}$ | $1.5×10^{-3}$ | 27.1 |

**Table S2** Graphics memory cost and computing parameter of different models for identifying discrete and aggregated vacancies in the 10 s irradiated samples ((For ResNet 18 model training, the patch size is 128×128, the batch size is 32).

| Model | Time cost (s) | Graphics memory cost (MB) | Computing Parameter (M) | FLOPs (M) |
|---|---|---|---|---|
| ResNet 18 | 2.9 | $1.5×10^{3}$ | 11.7 | $2.7×10^{7}$ |
| EGNN | 0.3 | $3.4×10^{2}$ | $1.5×10^{-3}$ | $3.3×10^{3}$ |
| Multiscale U-Net for atomic column localization | 0.2 | $1.5×10^{4}$ | 1.8 | $2.2×10^{6}$ |